\documentclass{aa}  
\usepackage{graphicx}
\usepackage{txfonts}
\usepackage{natbib}
\begin{document}
   \title{Metallicities \& Activities of Southern Stars}


   \author{J.S.~Jenkins\inst{1,2}
	  \and 
	  H.R.A.~Jones\inst{2}
	  \and 
	  Y.~Pavlenko\inst{2}
	  \and 
	  D.J.~Pinfield\inst{2}
	  \and 
	  J.R.~Barnes\inst{2}
	  \and 
	  Y.~Lyubchik\inst{2}
   }

   \offprints{J.S Jenkins}

   \institute{Department of Astronomy and Astrophysics, Pennsylvania State University, University Park, PA16802\\
     Center for Astrophysics, University of Hertfordshire, College Lane Campus, Hatfield, Hertfordshire, UK, AL10 9AB\\
              \email{jjenkins@astro.psu.edu}
	      \thanks{Based on observations made with the ESO telescopes at the La Silla Paranal observatory under program ID's 076.C-0578(B) and 077.C-0192(A)}
            }

   \date{Received September 2nd 2007}

 
  \abstract
   {}
   {We present the results from high-resolution spectroscopic measurements to determine metallicities and activities of bright stars in the southern hemisphere.}
   { We have measured 
the iron abundances ([Fe/H]) and chromospheric emission indices (log\emph{R}$'_{\rm{HK}}$) of 353 solar-type stars with
$V$=7.5$-$9.5.  [Fe/H] abundances are determined using a custom $\chi$$^{2}$ fitting
procedure within a large grid of Kurucz model atmospheres.  The chromospheric activities were determined by measuring the amount of emission in the cores of the
strong Ca\sc ii \rm HK lines.}
   {The sample of metallicities has been compared to other [Fe/H] determinations and was found to agree with these at the $\pm$0.05~dex level for spectroscopic values and at the 
$\pm$0.1~dex level for photometric values.  The distribution of chromospheric activities is found to be described by a bimodal distribution, agreeing well with the conclusions
from other works.  Also an analysis of Maunder Minimum
status was attempted and it was found that 6$\pm$4 stars in the sample could be in a Maunder minimum phase of their evolution and hence the Sun should only spend a few per 
cent of its main sequence lifetime in Maunder Minimum.
}
   {}

   \keywords{Stars: abundances --
                Stars: activity --
                Stars: atmospheres --
		Stars: planetary systems
               }

   \maketitle
%

\section{Introduction}

High resolution spectroscopy is an important tool used in the understanding of stars in the local neighbourhood
and can give an insight into various fundamental properties of the stars themselves (e.g. \citealp{edvardsson93};
\citealp{wright04}; \citealp{valenti05}).  Indeed, knowledge of stellar fundamental properties can also have a
direct bearing on other facets of stellar astrophysics such as the potential for discovering planetary systems.
After only three exoplanets had been discovered \citet{gonzalez} noticed all were orbiting stars with a higher
metal content than the Sun.  This trend has continued and been confirmed by various authors, most recently by
\citet{fischer05} who have shown that the probability of discovering an exoplanet around a solar-type star is
proportional to the stellar metal abundance following a power law described by 0.03x10$^{2.0\rm{[Fe/H]}}$, consistent 
with the predictions made by the core accretion models of planet formation.  Along with
a star's metallicity another important feature is the star's chromospheric activity.

Stellar chromospheric activity is intimately related to the stellar
dynamo, magnetism and the stellar rotation (\citealp{middelkoop82}; 
\citealp{middelkoop82b}; \citealp{rutten84}).
Since the measurement of activity gives us a feel for the motions on
both large and small scales in a stars atmosphere, it can be used to
select the most inactive and hence stable stars to scrutinize for
exoplanets.  It has been found that chromospheric activity can mask
or in some cases exhibit planetary radial-velocity signatures (e.g.
\citealp{queloz}; \citealp{henry02}).
Therefore it seems prudent for any radial-velocity planet search
program that is aiming to reach precisions of a few ms$^{-1}$ to
select the most inactive stars.  We have begun a radial-velocity
project to detect both hot Jupiter-type exoplanets and lower-mass
hot Saturns and Neptunes that have the potential to transit their
host star.  In Section 2 we discuss all observations and reduction
methods.  In Section 3 the methodology for extracting the spectroscopic information is explained: both the
activity analysis and the stellar photospheric properties such as metallicity abundance.  Section 4 compares
the values determined here with those of other works in the literature.  In Section 5 the results are discussed 
in terms of both the current status of exoplanets and stellar evolution.
Lastly, in Section 6 we discuss the conclusions of the project and plans for future work using this dataset.


\section{Observations \& Reduction}

All target stars and calibration data were observed using the Fibre-fed Extended Range Optical Spectrograph (FEROS)
mounted on the MPG/ESO - 2.2m telescope on the La Silla site in Chile.  To gain access to the entire
southern sky the run scheduling was split across two independent observing runs seven months
apart.  The runs were three nights in length, from the 2$^{\rm{nd}}$ to the 5$^{\rm{th}}$ of February and September 2006.  This 
culminated in over 350 stellar observations with exposure times in the range 120-480 seconds.
These exposure times gave rise to S/N ratios of 100-200 in the continuum around the iron line at 7500\AA\ and a median of $\sim$60 at the CaHK lines (3955\AA), 
with only a small tail of low S/N stars reaching down to values of 30. 
All calibration files needed for the reduction of the stellar spectra (flat-fields, bias and arc frames) were
obtained at the beginning and end of each nights observing, following the standard ESO calibration plan.

The reduction of all the spectra followed the standard reduction techniques.  Firstly the bias
signal was removed from each individual frame.  The overscan region was then subtracted from all frames individually and then trimmed off the image so as not to confuse any 
of the extraction algorithms.  FEROS
echelle data shows strong curvature in the 2D echelle image and, due to the use of an image slicer, has a complex slit cross-profile.
The curved echelle profile was straightened by fitting polynomials along the orders.
These were then precisely
traced using a well exposed Halogen flatfield image to follow the lamp light path.  Due to the half moon-like pattern of the sliced beams
the orders were smoothed in the spatial direction
to create a single smoothed order.  This allowed the tracing algorithm to properly follow the center of the order and not vary between the
two sliced peaks.  The traces
were then clipped to remove any stray pixel counts and to help tighten up the trace and then the object and background apertures were
selected.  To correct for the CCD pixel-to-pixel response we obtained 10+ flatfields with S/N ratios close to 1000 each morning and evening.  Each pixel was median
combined together to create a master flatfield frame.  The master flats were then used to calculate the balance factors by smoothing in
the dispersion direction and dividing
the smoothed flat by the master flat.  These balance factors were then multiplied into each science frame to flat field each image.
ThAr+Ne and ThArNe lamps were used to wavelength calibrate all the data.
The sky background was
negligible; however the scattered light was still removed.  This was done by setting the dekker limits large enough to include wide
inter-order spacing regions and then a low-order polynomial was
used to sample any gradients along the orders.  A second low-order polynomial was then used to fit to the total intensity over the whole
order to better model the scattered light.  The profile of each order was measured by sub-sampling each order profile
individually using Gaussians.  This
model was used to extract the object with an optimal extraction algorithm.  The data were then binned to
a linear wavelength step of 0.03\AA /pixel, or a mean velocity resolution of 1.42kms$^{-1}$/pixel, and shifted
into the rest frame by cross-correlating with an observation of HD102117 (G6V), which has been shown to exhibit a radial-velocity variation
$<$20ms$^{\rm{-1}}$ (\citealp{jones02b}).

\section{Methodology}

\subsection{Activities}

\begin{figure}
\vspace{4.0cm}
\hspace{-4.0cm}
\includegraphics{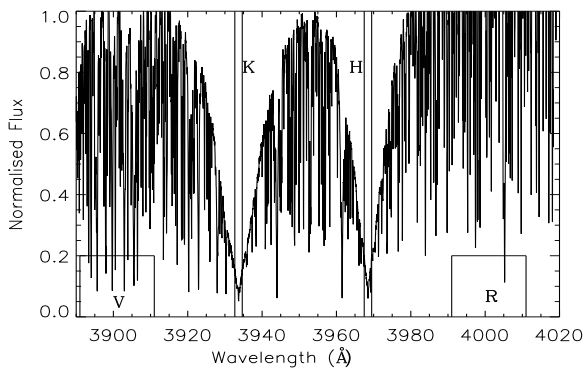}
\vspace{0.5cm}
\caption[FEROS spectra of HD59100]{The extracted FEROS spectra for the star HD59100 around the Ca\sc ii \rm K and H lines.  The star has
a spectral type of G2V and is chromospherically quiet (log\emph{R}$'_{\rm{HK}}$=-4.92).  All the
four passband regions (V,K,H\&R) have been highlighted on the plot.  The spectrum has been normalised to the continuum region
3991-4011\AA\, which represents the R passband.  All other features are relatively much weaker absorption lines, not noise.}
\label{feros_spectra}
\end{figure}

The methodology used to extract these high resolution activity indices is similar to that employed in previous works by the Anglo-Australian Planet Search 
e.g. \citet{tinney02}; \citet{jenkins06c} (hereafter J06) and Mount Wilson (\citealp{duncan}) projects.  Four passbands situated in a well concentrated region of the UV were 
used for flux
measurements.  Fig.~\ref{feros_spectra} shows the part of the spectrum used for all stars.  This particular spectrum is of HD59100, which
is a similar star to the Sun with a spectral type of G2V and has a final activity of log\emph{R}$'_{\rm{HK}}$~=~-4.92. All the four
passbands used to make the activity measurements have been shown for reference.  Each of the individual lines apparent in 
Fig.~\ref{feros_spectra} are relatively weaker spectral features, not noise.

\subsubsection{\emph{S}$_{\rm{MW}}$ Calibration}

\begin{figure}
\vspace{4.5cm}
\hspace{-4.0cm}
\includegraphics{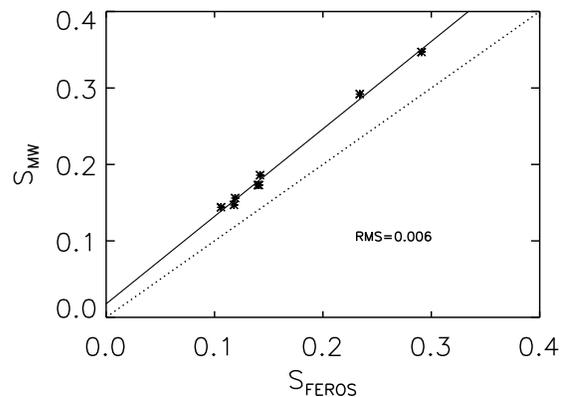}
\vspace{0.5cm}
\caption[\emph{S}-index calibration between FEROS and the Mt.~Wilson H\&K Project]{This plot shows the linear fit applied to the FEROS
spectra to convert the dataset to the MW
system of measurements.  All stars used as calibrators were taken from the MW project.  The fit shows an increasing linear trend with
increasing \emph{S}-index.  The overall
scatter of the points is highlighted on the plot, and with a value of 0.006, this is in agreement with the findings of \citet{tinney02},
\citet{wright04} and J06 who all calibrated using high resolution data and setups similar to the MW project.}
\label{s_cal_feros}
\end{figure}

\citet{tinney02}, J06 and \citet{wright04} have shown that linear calibrations are needed to calibrate high resolution
activity indices onto the MW system of measurements (\citealp{duncan}).  The setup employs four separate
passbands centered on and around the Ca\sc ii \rm HK lines.  The passbands used to estimate the mean flux in the continuum (V \& R) are
both square and centered at
wavelengths of 3891\AA\ and 4001\AA\ respectively.  Whereas the passbands used to obtain the mean flux in the K and H line cores have
triangular profiles with full-width half maximum's (FMHM) of 1.09\AA\ and are centered on the lines themselves at wavelengths of
3933.664\AA\ and 3968.470\AA\ respectively.

The following relation (Eq.~\ref{eq:feros}), is used to determine the FEROS activity indicator \emph{S}$_{\rm{FEROS}}$:

\begin{equation}
\label{eq:feros}
S_{\rm{FEROS}} = \frac{N_{\rm{H}} + N_{\rm{K}}}{N_{\rm{R}} + N_{\rm{V}}}
\end{equation}

Here the \emph{N$_{i}$} is the number of counts in each bandpass (where \emph{i}=H,~K,~V~and~R) and hence this provides a measure of the
ratio of the flux in the
Ca \sc ii \rm line cores to that of the continuum.  When this index was created for use at Mt.~Wilson it was tailored towards use with the
photometer and passband wheel
there and also employed a normalising factor, however by use of a similar, yet artificial setup, this procedure works very well for
extracting activity indices of bright solar-type stars.

A number of stars from \citet{duncan} were selected for use as calibrators onto the MW system, these are shown in Table.~\ref{tab:calibrators1}.  
For any activity study, the selected comparison stars should not introduce any offset into the final \emph{S}-indices, however this remains 
difficult due to the lack of long-term monitoring data in the literature for stars in the southern hemisphere.  Two of our calibration stars, 
HD1835 and HD10700, have recently had their long-term variability and activity trends published.  \citet{lockwood07} have shown that HD1835 
did exhibit variation at the level of $\sim$$\pm$0.05 between 1985 and 2002.  However, the star only spent a limited time at these extremes and was mostly 
found to be only $\sim$$\pm$0.02-0.03 from the mean activity value.  \citet{hall07} found a mean \emph{S}-index for this star of 0.362 using 82 observations 
over a period of $\sim$11~years, and this mean is 0.012 larger 
than the final computed value here.  HD10700 was also measured by Hall et al. and they found the mean \emph{S}-index to be 0.175 over 10~years and 67 observations, which is close to the 
value of 0.179 found here (i.e. two measurements of 0.178 and 0.180) within the uncertainties. 
Fig.~\ref{s_cal_feros} shows the linear
fit between the FEROS and the published MW values.  The fit has a gradient of 1.144$\pm$0.001 and an intercept of 0.018$\pm$0.006.  The
overall scatter of the points around
the linear trend is 0.006, which is similar to the scatter found by J06.  This low level of
scatter gives confidence in the robust nature of the reduction and analysis procedure.  The calibration error is found to be
consistent with the removal of any of the data points. 

\citet{wright04} and J06 used measurements of the stable star $\tau$~Ceti (HD10700) as a proxy for the random uncertainty
introduced in the reduction procedure.  As
$\tau$~Ceti has been shown to be extremely stable (\citealp{baliunas}) this is a useful tool for estimating the uncertainty introduced in
the reduction procedure.  Only two
measurements of this star were possible due to the star's position and bad weather, and hence it could only be used as a weak uncertainty
estimator.  The measurements were obtained
over two separate nights and have shown the
stability of the reduction.  The computed \emph{S}-index for both measurements are 0.178 and 0.180, indicating that the procedure is
robust.  As the reduction uncertainty is
weakly estimated here, the RMS scatter can be used as a first approximation of the total uncertainty, and also by comparison with the
findings from \citet{wright04} and J06 where both find uncertainties of $\sim$3-5\%.  We estimate that the total error 
included in each individual \emph{S} activity indicator is around 3-5\%.

\subsubsection{log\emph{R}$'_{\rm{HK}}$ Fit}

\begin{figure}
\vspace{4.5cm}
\hspace{-4.0cm}
\includegraphics{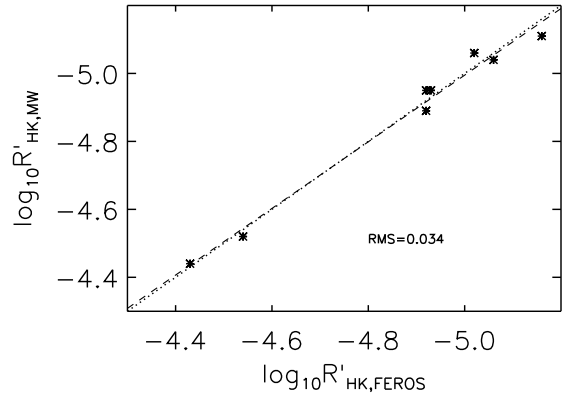}
\vspace{1.0cm}
\caption[log\emph{R}$'_{\rm{HK}}$ relation between FEROS and Mt.~Wilson]{The final log\emph{R}$'_{\rm{HK}}$ values for the FEROS
calibration dataset against the values
published in \citet{duncan} (MW).  There appears a tight fit around the 1:1 relation, which is shown by the dotted line.  The dashed line
represents the best-fit linear
trend to this data and the tight relationship means no further calibration is required.  The RMS scatter around the fit is shown in the
plot and is found to be 0.034.}
\label{rhk_fit_feros}
\end{figure}

The measured \emph{S}-index contains photospheric information and for activity analysis one wants exclusively chromospheric
information, therefore the methodology of \citet{noyes} has been used to remove the photospheric component of the activity index.
This method normalizes the flux to the bolometric luminosity of the star by use of empirical relations (see Noyes et al. for equations).
Fig.~\ref{rhk_fit_feros} shows the final log\emph{R}$'_{\rm{HK,FEROS}}$ values for all stars used to calibrate onto the MW system of
measurements.  The plot highlights
the close relationship between the published MW values from \citet{duncan} and these values.  A linear best-fit trend is applied
to the data and this
is shown by the solid line in the plot.  The gradient of this trend is 0.981$\pm$0.003 and has a negative intercept of -0.092$\pm$0.250.
The slope of this fit is
in agreement with that found by J06.  No other calibration is necessary.  The RMS
scatter here is 0.034, compared to $\sim$0.04 at the AAT.  Due to the number of uncertainties that can be introduced in the reduction
procedure, such as the
scattered light removal, blaze extraction, flatfielding etc, we believe that this RMS error is the best overall estimate of the errors
associated with each
individual activity measurement.  When we compare the final Hall et al. log\emph{R}$'_{\rm{HK}}$ values for the two calibration stars 
HD1835 and HD10700 against the values presented here, we see good agreement between the two within the uncertainties expressed i.e. 
-4.41/-4.43 and -4.94/-4.91 respectively.  All final log\emph{R}$'_{\rm{HK,FEROS}}$ indices, along with their associated \emph{S}-indices, 
Johnson colours and visual magnitudes from Hipparcos (\citealp{perryman}) are shown in Table~\ref{tab:activity1}.

\subsection{Metallicities}

When attempting to obtain accurate metallicity abundances ([M/H]) for solar-type stars some prior knowledge of the stellar envelope
conditions can be useful.  Various
important stellar parameters are input into spectral synthesis codes to model stellar spectra, these are the effective temperature
(T$_{\rm{EFF}}$),
the stellar surface gravity (log\emph{g}), the microturbulence parameter ($v_{\rm{mic}}$), the macroturbulent velocity ($v_{\rm{mac}}$)
and the stellar rotational velocity
($v_{\rm{rot}}$).  All of these parameters serve to either alter the line profiles and/or strengths of elemental abundances in the stellar
envelope and can have marked effects on any abundance determination.

\subsection{T$_{\rm{EFF}}$ Determination}

Various tables consisting of numerous stellar parameters and calibrations have
been published in the literature.  For example, Str{\"o}mgren \emph{ubvy}-photometry for a host of stars can be found in \citet{hauck} and
using calibrations from
\citet{olsen84} the effective temperatures and surface gravities can be found.  Also the catalogue of \citet{nordstrom04} has an extensive
list of effective temperatures
for over 16,000 stars.  \citet{nordstrom04} has also published the Str{\"o}mgren derived [Fe/H] values for all stars in their catalogue
and, along with the calibrations
of \citet{haywood} for solar-type stars in the \citet{hauck} catalogue, a large library of Str{\"o}mgren [Fe/H] can be compiled.  A large
number of stars in this study have measured Str{\"o}mgren indices.  Such measurements only allow for an estimate of the overall
metallicity abundance.  Spectral analysis is a more accurate
technique for the determination of the overall metallicity abundance, albeit at greater labour, as the information contained in each line
is measured individually and then
combined.  Whereas using photometry means losing all the information on each individual element and relying on the stability and accuracy
of the photometry and also the calibration of the scale with stellar parameters.  Hence, all stars with Str{\"o}mgren indices were kept in this observing program but priority 
was given to stars with no information at all.  Also as most stars in this project
did not have measured Str{\"o}mgren indices a different method was needed to obtain the stellar temperatures.

Recently a common method of temperature determination is by use of fitting the stellar spectra themselves (e.g., see \citealp{santos03};
\citealp{santos04};
\citealp{valenti05}).  However this method can be quite computer intensive and quantifying the errors can be quite difficult.  A simpler
method for obtaining the effective
temperature is by using photometric colour-effective temperature relations (e.g., \citealp{smith95})
(equation 8.9), which was recently employed by \citet{bond06} to determine the effective temperatures of stars on a full target list.
The Infrared Flux Method (IRFM) (\citealp{blackwell94}) was the calibration chosen to generate the effective temperatures using the Hipparcos $V$ magnitude and the 2MASS $K_{\rm{s}}$
magnitude of all these stars.  Since the 2MASS $K_{\rm{s}}$ is a shorter band measurement than the typical Johnson $K$-band, a magnitude correction taken from \citet{carpenter01} was 
used to correct the $K_{\rm{s}}$ magnitude for application to the IRFM.  The calibration from Blackwell \& Lynas-Gray was used to generate the final effective temperatures for all 
stars.  The $V-K$ 
colours have been found to be more reliable as a temperature indicator due to less metallicity independence, reduced line blanketing, the colour variation is less 
susceptible to surface gravity (\citealp{alonso96}) and the colours span a wider range of values compared to $B-V$ colours, giving a tighter sequence.

\subsubsection{Surface Gravity}

The surface gravity (log\emph{g}) of stellar objects is another parameter that can alter stellar line strengths, therefore altering the final abundances measured.  For each
of the stars in this list the $\chi$$^{2}$ for each chosen absorption lines was measured for a range of log\emph{g}'s, 3.5-5.0~dex in steps of 0.5.  This
allowed a wide parameter space to be probed and to test if the final metallicities were drastically affected by the surface gravity.  It was found from $\chi$$^{2}$ fitting
that the metallicity of these objects are only weakly dependent on the surface gravity and by interpolating between the points in $\chi$$^{2}$-space the error introduced is
minimal.  

However, the analysis in Section~5.3 requires some knowledge of the stellar surface gravity to use as a proxy for evolutionary status and this was done by searching the stellar 
spectra for a large region that would produce significant enough gravity sensitivity within our course grid of gravity models.  The region of spectra chosen was centered at 6200\AA\ 
and had a width of 200\AA.  By extracting this region from all stellar spectra and comparing it to the same region extracted from the models, for each of the gravity values and best 
fit temperatures and metallicities, then the $\chi$$^{2}$ minimisation could be used to indicate the evolutionary status of the star.  Since this was across a course grid in surface 
gravity space only an indication could be made, and hence construction of a grid with higher resolution would be required to determine the actual surface gravity with any degree of 
certainty.  However, a check of a number of objects that overlap \citet{valenti05} and
\citet{gray05} have shown that the $\chi$$^{2}$ can determine the minimum and a selection can be performed to remove any low surface gravity objects.  For instance, the star
HD27442 has been shown to have a log(\emph{g}) of 3.78 by Valenti \& Fischer and 2.93 in Gray et al. and the $\chi$$^{2}$ here minimises at 3.5, which is the lowest gravity
value in this study.  This shows the technique can be used to select against too evolved and/or too young stars. 

\subsubsection{Microturbulent Velocity}

Another parameter that has been introduced to smooth over the differences obtained between the predicted and observed equivalent widths of absorption lines as a function of
line strength is the microturbulence parameter ($v_{\rm{mic}}$).  This arbitrary parameter is not necessarily required in 3D convective models of stellar
atmospheres (\citealp{asplund00}) but has become a standard when dealing with one-dimensional stellar spectra to alleviate any discrepancies in equivalent
width measurements.  $v_{\rm{mic}}$ is used to represent uncertainties in a number of parameters like excitation energy, oscillator strength, effective temperature, surface
gravity etc.  Valenti et al. find that $v_{\rm{mic}}$ and the total stellar abundance ([M/H]) are partially degenerate therefore they fixed the microturbulence
velocity to 0.85kms$^{-1}$ to minimise errors in their final abundance values. In this study the microturbulence parameter has already been fixed to 2kms$^{-1}$ as the
\citet{kurucz93} model atmospheres were computed for this value across all the required metallicity range and also to create a consistent sample.  This does not introduce any 
significant uncertainty since our line list selection 
aimed to target weak lines that are formed deeper in the stellar interior and hence occupy the linear part of the curve of growth.  Such lines are 
insensitive to changes in the microturbulent velocity.  Hence, we believe that the final $\chi$$^{2}$ estimated error represents the uncertainty of this parameter.  To fully test this 
we computed 10 values using microturbulences of both 1km/s and 2km/s for solar-type stars with a range of metallicities and found no offset of the final abundances, with only a 
small RMS scatter $\pm$0.02~dex.

\subsubsection{Macroturbulent Velocity \& Stellar Rotation}

In the first approach the macroturbulent velocity ($v_{\rm{mac}}$) parameter has the same affect on atomic line profiles as that of rotation or the instrumental profile, 
however it is described by a different profile than both these parameters.  As it's name suggest, $v_{\rm{mac}}$ is the term that considers
the smaller scale motions of the stellar envelope induced by internal motions within the star.  High-resolution spectra and detailed imaging of the Sun's surface that considers the 
larger scale motions has revealed
velocity fields that vary typically at a few kms$^{-1}$ (e.g. \citealp{rieutord01}).  These motions are generated by photospheric granulation and acoustic waves permeating
throughout the star due to subsurface convection.  This motion is also present in the high-resolution stellar spectra generated from the integrated flux from a stellar
source and manifests itself by broadening line profiles.  However, as the average resolution across the spectrum is $\sim$0.15\AA, (a velocity of $\sim$6kms$^{-1}$), we 
believe the broadening to be dominated by the stellar rotation as velocity fields of a few kms$^{-1}$ are comparable only with the slowest rotation rates.

As mentioned above, stellar rotation ($v$sin\emph{i}) has the effect of broadening line profiles without altering the line equivalent width, therefore knowledge of the rotational
velocity allows a better representation of the line profile and will further reduce the errors on the final metallicities.  In order to determine the $v$sin\emph{i} for each
star we selected an iron line that was shown to exhibit little change in line strength through the range of metallicities.  This line, which is centered at a wavelength of
7445.758\AA, is highlighted by an asterisk in Table~\ref{tab:line_list} and is the strongest line in the list.  The observed spectra 
around this line were then compared with the model spectra, broadened in steps of 1kms$^{-1}$ between 1-15kms$^{-1}$.  We chose a width of $\pm$0.5\AA\ centered on the line 
to compare each line with 
the models as this also included valuable wing information.  $\chi^{2}$ values were generated at each velocity step and the minimum represents the final $v$sin\emph{i}.  The
resolution of the instrument at 7000\AA\ is $\sim$6kms$^{-1}$, however we found that we can discern values of $v$sin\emph{i}$>$~2.5kms$^{-1}$.

\subsubsection{Model Atmospheres}

To enable highly accurate chemical abundances to be determined from this dataset a number of utilities were used to generate relevant
synthetic spectra to compare with the observations.  The program chosen to create the synthetic spectra was WITA6 (\citealp{pavlenko00}).  
\citet{moore56} was used for the atomic line list and damping constants, with the ATLAS 9 model atmospheres of \citet{kurucz93} and oscillator strengths 
from \citet{gurtovenko98}.

The grid of Kurucz models range from 3500K to 8000K with surface gravities and metallicities from
3.5 to 5.0 and -1.5 to 1.0 respectively, ensuring
coverage of almost all possible values for these parameters for all objects.  A few objects require lower metallicity models to
accurately measure their values.  A parameter that is fixed throughout is the microturbulent
velocity. This is held fixed at 2kms$^{-1}$ for all the model grids, which will introduce some error into the accuracy of the final
synthetic spectrum but after investigation it was found to be small compared with other uncertainties. 

Since most of our objects are G and K type stars these models are adequate for metallicity computation, however \citet{smalley97} have shown that 
there are errors inherent in them due to an inadequate convective overshooting algorithm.  Although the errors are small for G and K stars, they can be 
significant for hotter F type stars (\citealt{gray03}).  To investigate this we compute the metallicities for all objects with effective 
temperatures greater than 6000K, with and without convective overshooting.  The results from the convective overshooting On-Off gave a mean of -0.01$\pm$0.06~dex for 
a total of 19 data points, which is consistent with an additional scatter of 0.06~dex for the hottest stars in this sample. 

\subsubsection{Line Selection}

\begin{figure}
\vspace{0cm}
\hspace{-4.0cm}
\includegraphics{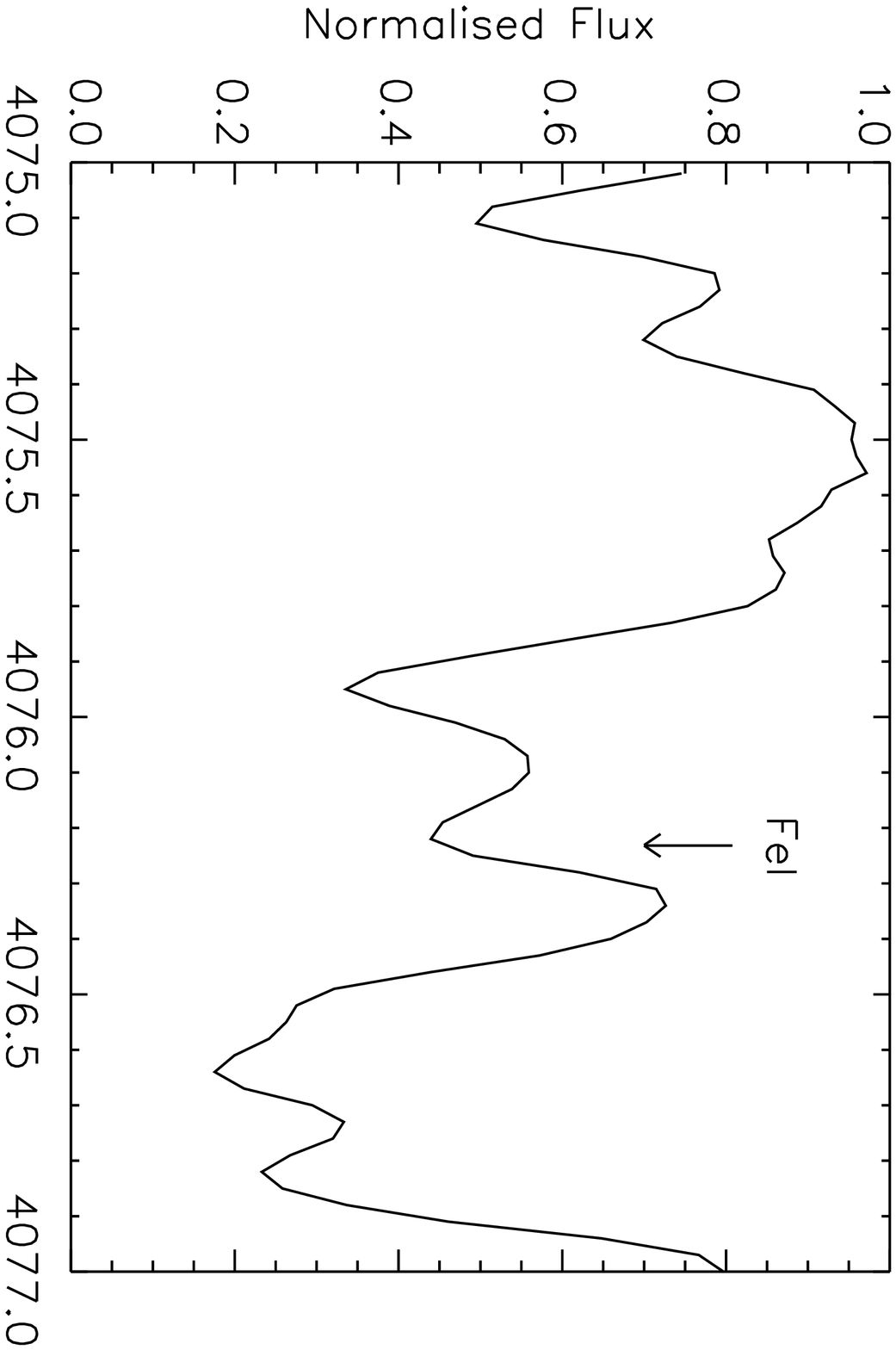}
\vspace{5cm}
\includegraphics{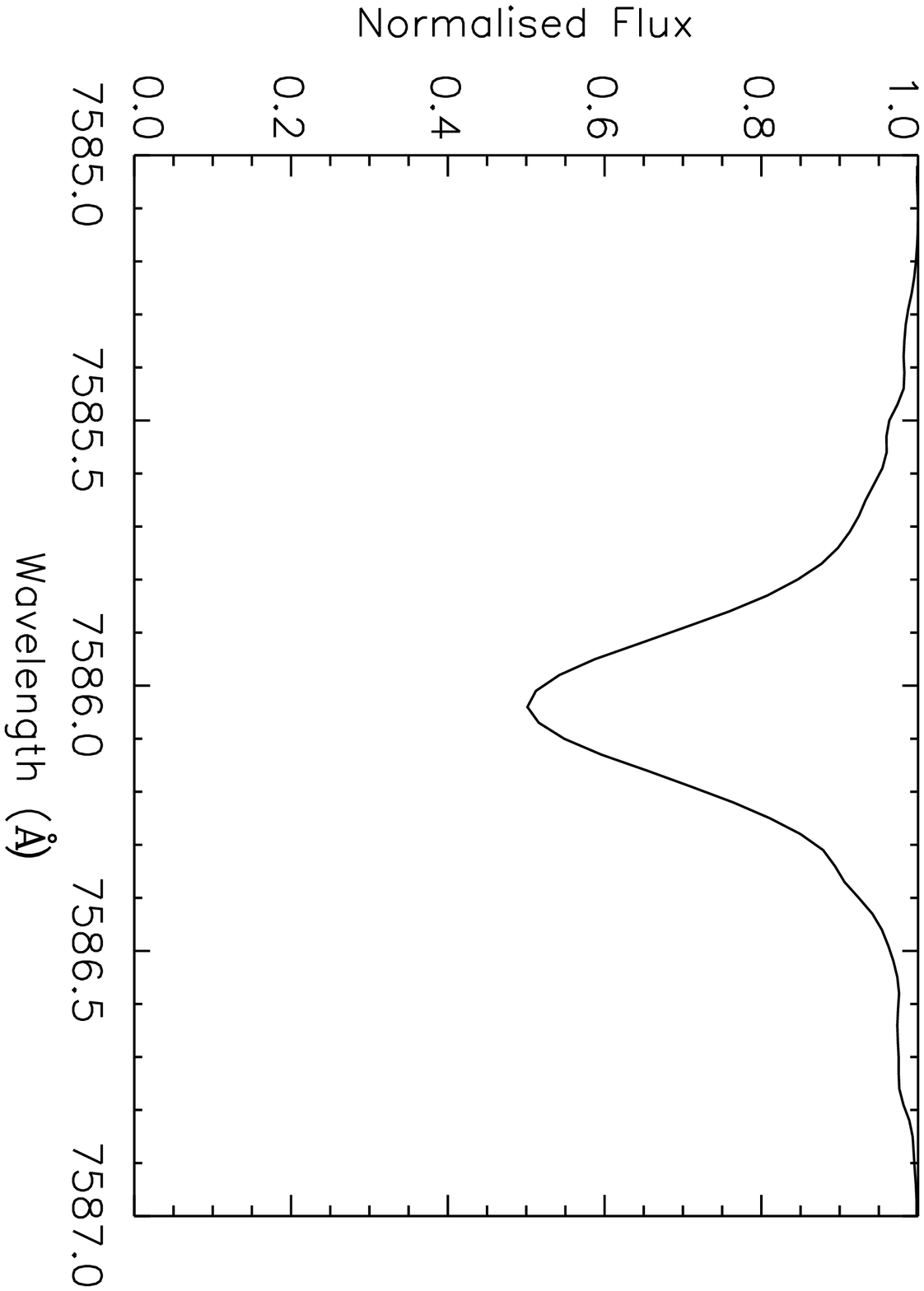}
\vspace{3.9cm}
\caption[Blended and unblended iron lines in the HD38459 spectrum]{A spectrum of the star HD38459 towards the bluer end of the stars
visual band is shown in the upper panel.
All of the lines in this wavelength region (4074-4078\AA) are blended.  The FeI line in the middle of this region (4076.22\AA) is clearly
blended at both sides.  The lower
panel shows the iron line at 7586.027\AA.  No strong line blends can be discerned from the noise at this spectral resolution and if any do exist they are rather 
marginal.}
\label{blended}
\end{figure}

One of the most critical parts of abundance determination is the selection of lines to use in the observed spectrum to obtain a clean line
profile.  A vast number
of the lines in a star such as the Sun are blended, especially towards the
bluer end of the visual band
(see Fig.~\ref{blended} upper panel).  The lines used in this work have been selected because they appear unblended at this
spectral resolution (Fig.~\ref{blended} lower panel) and they are also weak in the solar spectrum.  Weak lines lie on the linear part of the curve of growth, 
therefore they are more sensitive to large
changes in the model parameters and are better indicators of line profile characteristics giving more accurate
abundances.  A line's position on the curve of growth will significantly affect all parameters associated with the line
as the curve of growth determines how the line evolves when more absorbers are introduced.  For example, optically thick lines will saturate more quickly and therefore are not 
good estimators of the line strength and hence abundance status of the star.  Strong lines also add extra uncertainty through saturation, microturbulence and 
uncertainties associated with damping constants (e.g. see \citealp{gray05}).  

The FEROS spectra used here have a S/N and resolution sufficiently high enough to allow a large enough number
of apparently unblended lines to be used in this analysis.  The lower wavelength cutoff chosen was at a $\lambda~>~5400$\AA\ this was due to the increased blending 
factor mentioned above (Fig.~\ref{blended}), however
any lines below 6500\AA\ were very carefully selected to ensure they were isolated from other possible blends.  The
wide coverage region obtained using FEROS ($\sim$3500~-~9200\AA\ \citealp{kaufer99}) allows one to select a large enough number of unblended
iron lines.  Solar spectra from FEROS, taken as part of the standard ESO Calibration plan, was used as a
template to select lines that
would appear in all other spectra.  Also, no telluric lines were seen to blend with these lines.  All lines used for abundance determination are shown in Table.~\ref{tab:line_list}. 

\begin{table*}
\center
\caption[Iron lines used in the abundance determination]{Fe lines used in the abundance measurements are tabulated with their wavelengths, oscillator strengths and excitation
energies.  The iron line highlighted by the asterisk was used in our determination of $v$~sin~\emph{i}.} \label{tab:line_list}
\
\begin{tabular}{ccccc}
\hline
\multicolumn{1}{c}{Element} & \multicolumn{1}{c}{$\lambda$~(\AA)} & \multicolumn{1}{c}{\emph{gf}} & \multicolumn{1}{c}{E~(keV)} \\ \hline

Fe\sc i \rm & 5806.723 & 1.096E-01 &  4.610 \\
Fe\sc i \rm & 5852.215 & 5.623E-02 &  4.550 \\
Fe\sc i \rm & 5855.082 & 2.239E-02 &  4.610 \\
Fe\sc i \rm & 5856.086 & 2.291E-02 &  4.290 \\
Fe\sc i \rm & 6027.051 & 5.888E-02 &  4.070 \\
Fe\sc i \rm & 6151.621 & 4.266E-04 &  2.180 \\
Fe\sc i \rm & 6159.379 & 1.148E-02 &  4.610 \\
Fe\sc i \rm & 6165.359 & 2.754E-02 &  4.140 \\
Fe\sc i \rm & 6173.340 & 1.259E-03 &  2.220 \\
Fe\sc i \rm & 6229.227 & 9.120E-04 &  2.840 \\
Fe\sc i \rm & 6608.027 & 9.120E-05 &  2.280 \\
Fe\sc i \rm & 6627.547 & 2.570E-02 &  4.550 \\
Fe\sc i \rm & 6703.566 & 8.318E-04 &  2.760 \\
Fe\sc i \rm & 6725.359 & 5.370E-03 &  4.100 \\
Fe\sc i \rm & 6750.152 & 2.344E-03 &  2.420 \\
Fe\sc i \rm & 6810.266 & 8.710E-02 &  4.610 \\
Fe\sc i \rm & 7158.477 & 1.148E-03 &  3.650 \\
Fe\sc i \rm & 7189.152 & 1.585E-03 &  3.070 \\
Fe\sc i \rm$^*$ & 7445.758 & 6.761E-01 &  4.260 \\
Fe\sc i \rm & 7491.656 & 8.710E-02 &  4.300 \\
Fe\sc i \rm & 7568.906 & 1.288E-01 &  4.280 \\
Fe\sc i \rm & 7586.023 & 7.079E-01 &  4.310 \\
Fe\sc i \rm & 7751.109 & 1.514E-01 &  4.990 \\
Fe\sc i \rm & 7780.562 & 1.148E+00 &  4.470 \\
Fe\sc i \rm & 7807.914 & 2.754E-01 &  4.990 \\
Fe\sc i \rm & 7832.207 & 1.259E+00 &  4.430 \\
Fe\sc i \rm & 8047.625 & 2.239E-05 &  0.860 \\
Fe\sc i \rm & 8239.137 & 3.715E-04 &  2.420 \\

\hline
\end{tabular}
\medskip
\end{table*}


\subsubsection{Oscillator Strengths}

One of the larger sources of error comes from uncertainties in the oscillator strength values.  The oscillator
strengths chosen were extracted from the list of \citet{gurtovenko98}.  Gurtovenko \& Kostik measure the oscillator strength (\emph{gf})
using measurements of the equivalent widths, central intensities along with
the synthesis of the whole Fraunhofer spectrum of the specific chemical element.  They define two separate measurements of \emph{gf} using
separately the equivalent widths
of the lines and the central line depths to allow them to estimate the accuracy of the values and to generate estimates of the
uncertainties in the data.  They found that
the internal accuracy of the final log(\emph{gf}) values to typically be $\pm$0.07~dex.  Recently \citet{bigot06} have determined the
oscillator strengths for Fe\sc I \rm and Si\sc I \rm in
the Gaia spectral window ($\sim$8480-8750\AA) and found accuracies of typically $<$~0.1~dex with laboratory results, whereas the
oscillator strength accuracies found in databases such as the Vienna Atomic Line Database can be much lower.

\subsubsection{WITA Program}

The latest version in the WITA series of programs was used to generate the grid of synthetic spectra.  The WITA series were developed as part
of the ABEL6
complex (\citealp{pavlenko91}), which was composed partly from \citet{kurucz79} ATLAS subprograms.  The WITA6 program (\citealp{pavlenko00})
itself is a modified
version of the WITA2 program (\citealp{pavlenko95}).  All computations performed by WITA6 are employed in the classical way, using Local
Thermodynamical
Equilibrium (LTE) and assuming hydrodynamic equilibrium for one-dimensional model atmospheres.  The code allows for the extrapolation of
the model atmosphere.  The ionisation-dissociation
equilibria equations system has been solved for various temperature structures assuming LTE.  A Voigt profile was adopted for single
absorption lines, taking into
consideration all line broadening sources, such as natural broadening, thermal broadening, pressure broadening and resonance broadening.

\subsubsection{Continuum Normalisation}

No flux calibration was performed which left behind a gradient in
the spectral profile across the
measured wavelength region.  To attempt to detect the true continuum level around each line a cubic-spline fit procedure was performed
using the program \sc continuum \rm part
of the \sc iraf \rm library of routines.  A global fit was used across the
entire wavelength region.  By
employing tight constraints on the upper and lower limit rejection thresholds for the fitted data points a good continuum measurement was
made for the entire spectra. 

\subsection{$\chi$$^{2}$ Fitting}

\begin{figure}
\vspace{3.5cm}
\hspace{-4.0cm}
\includegraphics{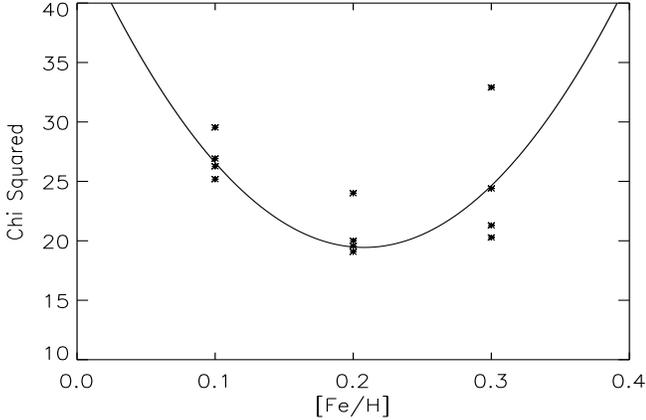}
\vspace{2.5cm}
\caption[$\chi$$^{2}$ parabola fit]{The $\chi$$^{2}$ fit for HD1835 in metallicity space.  The points represent the means of four $\chi$$^{2}$ values for different surface
gravities at each of these metallicities.  The solid line is the second-order polynomial fit to the points, with the final metallicity representing the minimum of this
function.}
\label{chi_fit}
\end{figure}

Once all the parameters for each object have been fit and iron lines selected the final step is to
generate the desired [Fe/H] abundance.  An IDL code has been developed
that will employ several different steps to compare the lines in the real spectrum with the corresponding lines in the
synthetic spectrum and generate a $\chi$$^{2}$ value for each line in each star.  These $\chi$$^{2}$'s will minimise
when the synthetic spectrum best replicates the real spectrum, allowing [Fe/H] abundances to be determined.

The first step in the program takes the generated temperature values for each star and then selects the spectra that have the nearest
temperatures above and below the generated temperature in the grid.  The Kurucz grid has temperature steps of 250K,
therefore to better replicate the real spectra the upper and lower spectra are interpolated to an improved temperature
scale.  
It is at this stage when the synthetic spectrum is broadened for
stellar rotation and the instrumental profile.  The rotation rates are generated following the procedure described above.
The instrumental profile is modeled by convolution with a Gaussian profile of width determined by the resolution of FEROS.
The FEROS resolution is $\sim$48'000, which at these wavelengths corresponds to a Gaussian of width $\sim$0.15\AA\, however since the instrumental profile changes with wavelength (e.g. 
telluric lines in the far red of the chip have widths $>$0.2\AA) we also use the models to better define an optimum broadening width.

\begin{equation}
\label{eq:chi}
\chi^{2} = \frac{(O^2-M^2)}{\sigma^2}
\end{equation}

After the model spectra have been prepared for comparison with the real spectrum the program selects each
individual line that has been selected for comparison and employs a $\chi$$^{2}$ fit between the real and model spectra.  Equation.~\ref{eq:chi} shows the equation employed
for this fitting procedure, where $O$ is the observed spectra, $M$ is the modeled spectra and $\sigma$ is the total errors in both.  This $\chi$$^{2}$ equation is applied
to each line by selecting the line center and comparing a region $\pm$0.2\AA\ from the center.  The 0.4\AA\
window was selected by eye after scrutinising all lines and finding the optimum window for these lines in the FEROS
spectra.  As this window is only an optimal window and will include more continuum in some lines than others, the median of all the $\chi$$^{2}$ values is used.  This helps remove any outlying values due to bad lines, any possible
small edge blending and extraction errors.  The median is generated for each metallicity value four times, representing
the four different log(\emph{g})'s for each metallicity.  Figure~\ref{chi_fit} shows the fitting output for the star
HD1835.  The crosses are the medians of the four separate log(\emph{g}) fits and the solid line represents the bestfit
second-order polynomial to the points.  The minimum of this function represents the stars [Fe/H] abundance, which for
this star is 0.21$\pm$0.03, agreeing with the \citet{valenti05} value of 0.25$\pm$0.03.  This entire procedure
is applied to all targets.  On a standard Pentium 4 processor with 1~Gigabyte RAM, it takes
around 2~days to complete.

\subsubsection{$\chi$$^{2}$ Uncertainties}

\begin{figure}
\vspace{4.5cm}
\hspace{-4.0cm}
\includegraphics{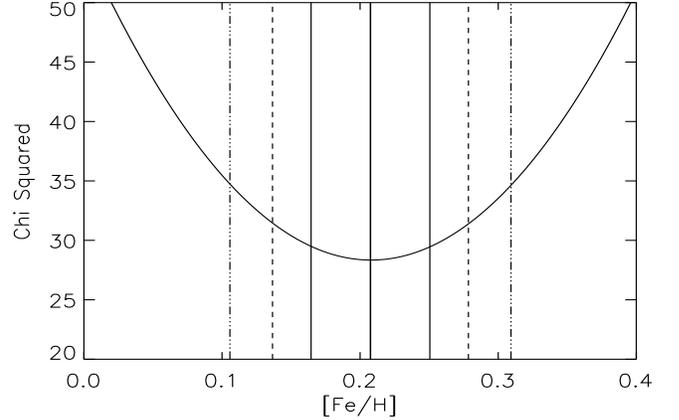}
\vspace{1.5cm}
\caption[$\chi$$^{2}$ error for the star HD1835]{The $\chi$$^{2}$ fit across all metallicity values for the star HD1835 with the 1$\sigma$ (solid bounds), 2$\sigma$ (dashed bounds) and
3$\sigma$ (dot-dashed bounds) error estimates shown.  These errors ranges were selected by interpolating a table of $\chi$$^{2}$ steps for a two parameter fit i.e. surface
gravity and metallicity.  The central solid line marks the minimum of the fitted function and represents the final abundance value derived for this star.}
\label{chi_errors}
\end{figure}

To determine the errors in the final abundance measurements the fitted $\chi$$^{2}$ parabola was used to interpolate the 1,2 and 3$\sigma$ values using a table of $\chi$$^{2}$
values.  Figure~\ref{chi_errors} shows the $\chi$$^{2}$ space for all three of these error ranges.  The solid lines bound the 1$\sigma$ error region, with the central solid
line marking the minimum of the function or the final abundance measurement.  The dashed and dot-dashed regions mark the 2 and 3$\sigma$ parameter space.  Since the function
is fit to both the log(\emph{g}) and the iron abundance a two parameter $\chi$$^{2}$ error step was performed.  The 1$\sigma$ errors have a step of 1.155 from
the minimum value, with the 2 and 3$\sigma$ steps equaling 3.118 and 6.396 respectively.  This error analysis is only valid assuming one has a handle on the uncertainty,
however as shown before their are a number of uncertainty sources that are difficult to quantify, (i.e. atomic oscillator strengths), therefore the final $\chi$$^{2}$
curve must represent this uncertainty.  To accomplish this the reduced Chi-squared ($\chi$$^{2}_{r}$) was also determined by dividing the $\chi$$^{2}$ by the number of degrees
of freedom.  The degrees of freedom were found by subtracting the number of fitted parameters from the number of data points for each iron line.  Ideally the
$\chi$$^{2}_{r}$ should equal, or be close to 1 for a good fit with all uncertainty sources well modeled.  However, since many uncertainties are poorly constrained, the 
$\chi$$^{2}$ function was normalised by the difference between the minimum of the $\chi$$^{2}_{r}$ function and the optimum (i.e. 1).  This has the same effect as randomly 
adding percentage errors into the original $\chi$$^{2}$ equation (Eq.~\ref{eq:chi}) to broaden the fitting function and increase the final uncertainty estimates for all 
abundances.

\section{Comparison With Previous Work}

\subsection{Activities}

\begin{figure}
\vspace{4.5cm}
\hspace{-4.0cm}
\includegraphics{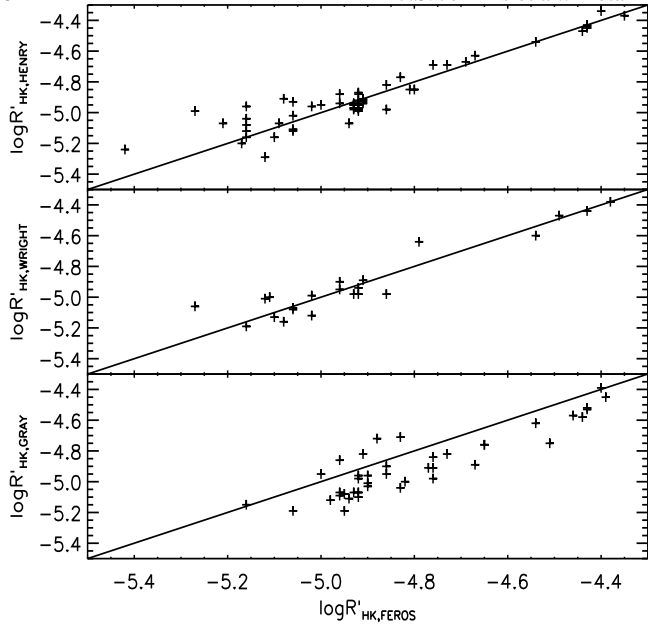}
\vspace{3.2cm}
\caption[Activity comparisons between this work and that from Henry et al. (1996), Wright et al. (2004) and Gray et al. (2006)]{A direct comparison of
log\emph{R}$'_{\rm{HK}}$ values from this work against \citet{henry} (upper), \citet{wright04} (middle) and \citet{gray06} (lower).  The plots show good agreement with
Henry et al. and Wright et al. and both have RMS scatters of $\sim$0.08 in log space.  However, there appears a small offset with the results of Gray et al., with the values
here slightly higher by around 0.1~dex.  The solid line through the data represent a 1:1 relationship.}
\label{comparison_plots}
\end{figure}

To make judgments and comparisons of this data to other published work we have plotted the individual activity values against those of three other catalogues.
Figure~\ref{comparison_plots} shows the comparisons of these values to those of \citet{henry} (upper panel), \citet{wright04} (middle panel) and
\citet{gray06} (lower panel).  The comparison to that of the Henry et al. values shows a scatter around the 1:1 relation (solid line) with no apparent offsets.  Considering
they also used calibrators from \citet{duncan} one would expect there to be a good relationship.  However, the Gray et al. work makes use of low-resolution
spectra and also uses a different setup from that employed here.  They centered 4\AA\ square bandpasses on the Ca\sc ii \rm HK line cores instead of triangular bandpasses of
width 1.09\AA\, which is used at Mt.~Wilson and also used here.  This will contribute to an increase in the scatter between the data and also to non-linear calibrations onto
the MW system.  The RMS scatter in this data is found to be $\sim$0.08.  A Student-T test was performed on all three of these distributions to test if they have the same
true means.  The test revealed an 83\% probability that the data from here and that of Henry et al. are drawn from the same mean, which given the intrinsic variability
associated with cool dwarf stars and the limited number of data points is noteworthy.

The middle panel in Fig.~\ref{comparison_plots} shows the data from \citet{wright04} against the values found here.  There appears a good fit to this data around the 1:1
relation with only two outliers.  This is a nice fit when considering the intrinsic scatter inherent in the values themselves, which can alter by 0.15 in
log\emph{R}$'_{\rm{HK}}$ over a period of a few years.  The RMS scatter here is $\sim$0.08, the same as the scatter of Henry et al.  However, the two outlying stars (HD8038
and HD150437) cause this unweighted RMS to deviate from the true scatter.  Once these two stars have been removed the RMS scatter is found to be 0.06, highlighting the close
relationship that is found using high-resolution spectra and a setup similar to that of the MW project.  This is highlighted by the fact that the T test for this data
reveal that they are both drawn from the same mean at a probability of 88\%.  Again this is high given the limited amount of data points to compare against.

\begin{figure}
\vspace{4.5cm}
\hspace{-4.0cm}
\includegraphics{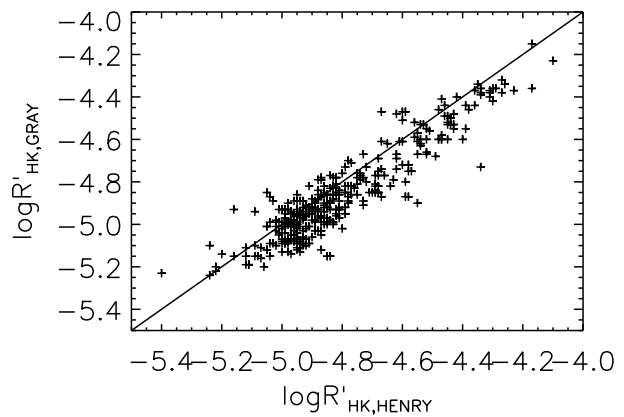}
\vspace{0.6cm}
\caption[Activity comparison between the works of Henry et al. (1996) and Gray et al. (2006)]{A comparison of the log\emph{R}$'_{\rm{HK}}$ activity indices from \citet{henry}
and \citet{gray06}.  There is an offset in the data, with the values of Gray et al. around 0.1~dex lower than those of Henry et al.}
\label{rhk_gray_henry}
\end{figure}

A recent large survey in the southern hemisphere has been conducted by \citet{gray06} connected to the NStars Project.  The lower panel in Fig.~\ref{comparison_plots}
 compares our measurements to this data.  There is some scatter around the 1:1 relation, however the majority of objects lie below this fit indicating a possible systematic
offset in the
data.  Neglecting the stars above the 1:1 fit there is an offset of around -0.1 in log\emph{R}$'_{\rm{HK}}$ that is not present in both other studies. A straight line fit 
to describe the offset has a gradient of 0.938$\pm$0.006 and offset of -0.389$\pm$0.362, with an RMS scatter of 0.096.  This can be used to transform these results to the scale of 
Gray et al.  The significance 
of the offset is at the 2$\sigma$ level as the T test for this comparison gave a probability of 96\% that both distributions were drawn from different true means.  We also performed 
the log\emph{R}$'_{\rm{HK}}$ conversion including all the stars in common with Gray et al., however once this calibration has been performed, both Henry et al. and Wright are 
offset from this work by around +0.1~dex.  The point can
be highlighted by plotting Gray et al. against the work of Henry et al. as both studies used similar resolutions and setups (Fig.~\ref{rhk_gray_henry}).
It can be seen from this comparison that the Gray et al. dataset is around 0.1~dex lower in log\emph{R}$'_{\rm{HK}}$, which is difficult to explain by intrinsic scatter.  
Indeed, the T test revealed a probability of 98\% that both have been drawn from different true means.
The offsets could arise due to the different calibrators used by this and the Henry et al. study against those of Gray et al.  Gray et al. used calibrating stars from
\citet{baliunas} compared to the Duncan et al. values, however as no offset appears between this study and Wright, (who also calibrated from Baliunas et al.), this seems
unlikely.  It may simply arise from a combination of the calibrators and the wide bandpasses used to extract the activities as the width, shape and resolution in the line
cores can serve to mask any emission feature and hence mask the true activity of a star.  Note also that Henry et al. performed non-linear calibrations onto the MW system, 
whereas both this work and Gray et al. use linear calibrations and this may explain some of the offsets seen, especially towards the more active end of the calibration.  
Whatever the cause it is clear a more thorough look at all these studies
together, along with long term variability data is needed to extract a good list of intrinsically stable stars to use as calibrators for similar studies in the southern
hemisphere.

\subsection{Metallicity}

\begin{figure*}
\vspace{4.5cm}
\hspace{-4.0cm}
\includegraphics{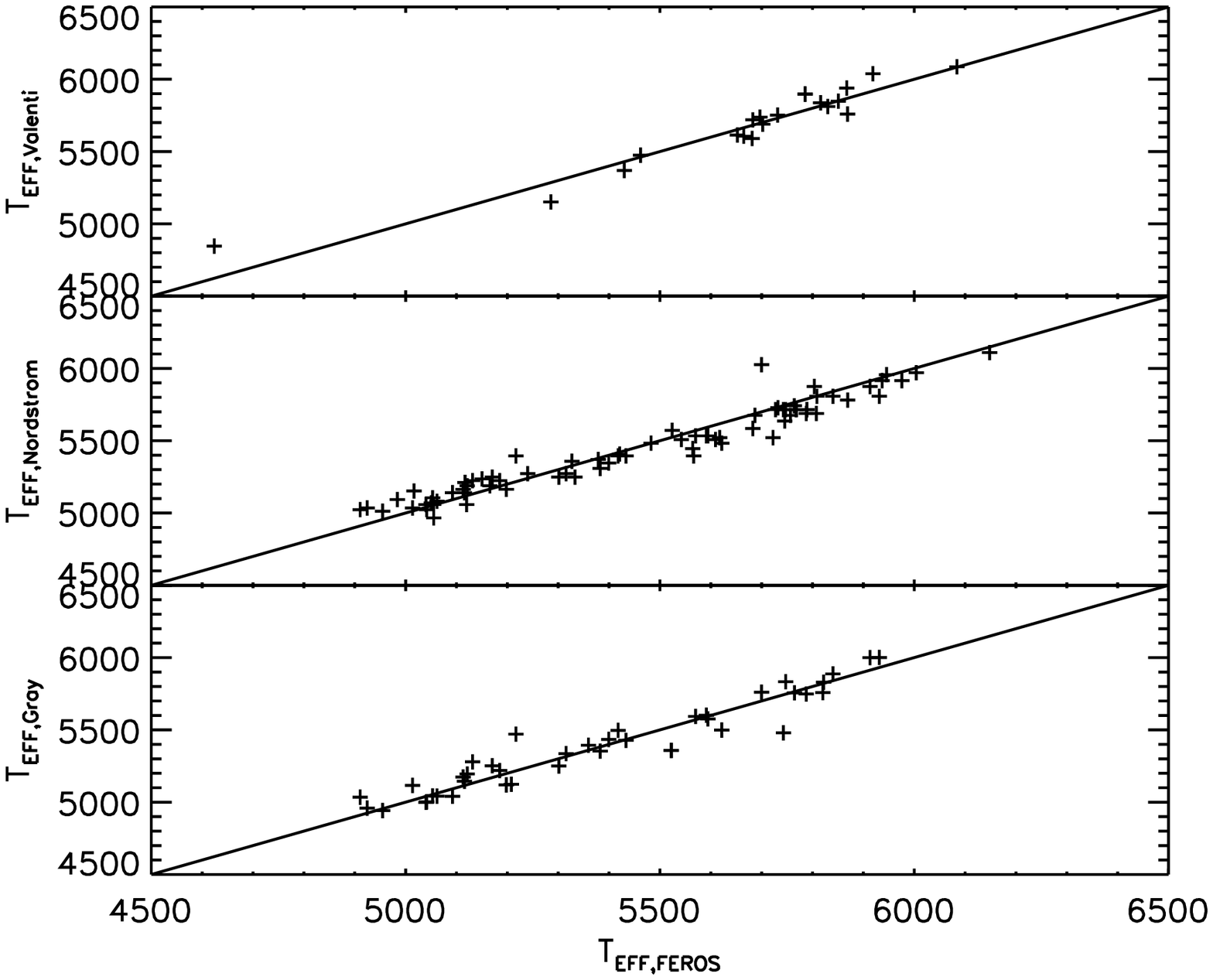}
\includegraphics{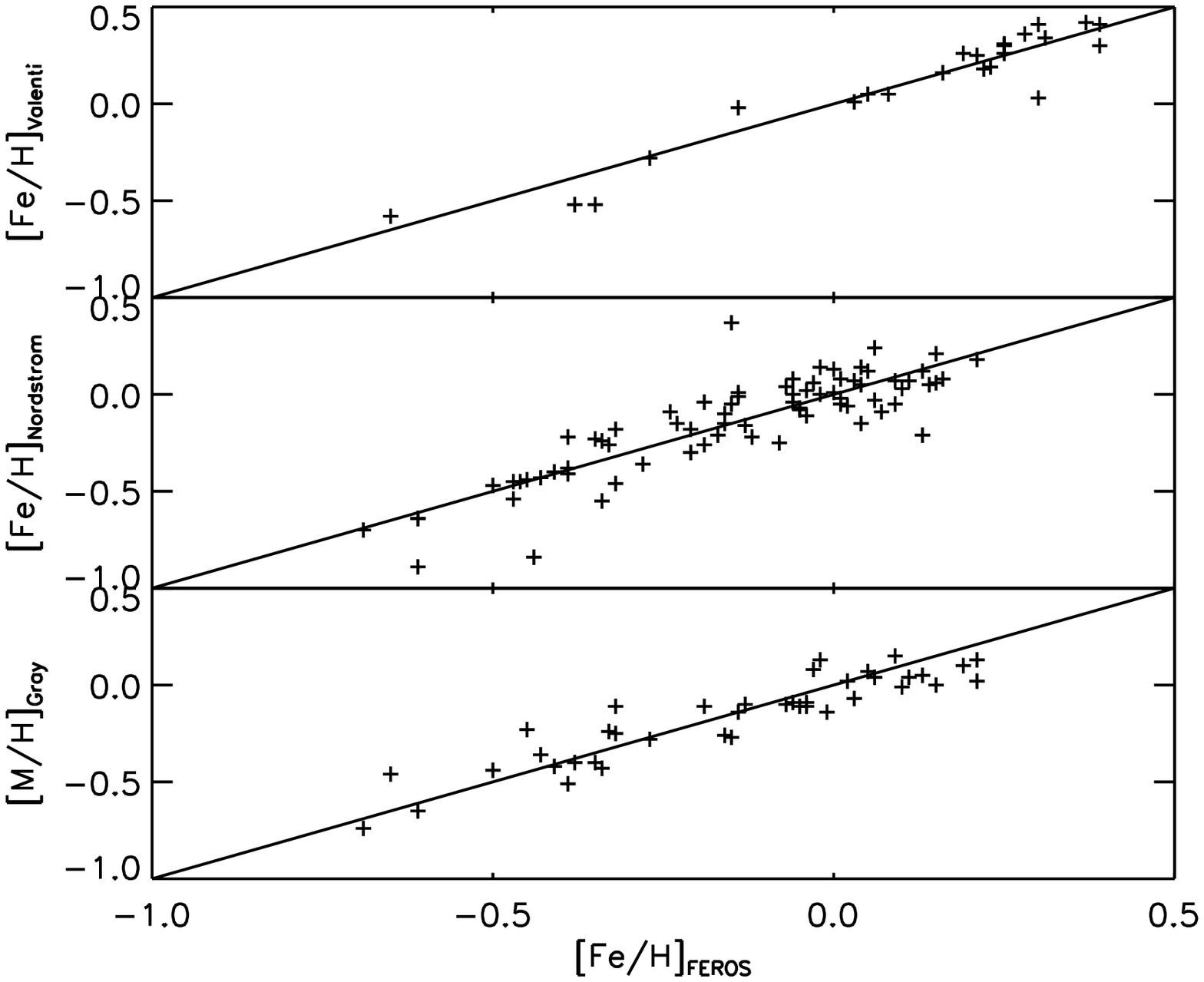}
\vspace{2.5cm}
\caption[Effective temperature and metallicity comparison between this work and Valenti \& Fischer (2005), Nordstr{\"o}m et al. (2004) and Gray et al. (2006)]{A direct
comparison of effective temperatures (left panels) and metallicity values (right panels)
determined here and those published in recent work.  The upper plots are a comparison with the values published in
\citet{valenti05}, which were derived using spectral line fitting similar to the analysis in this project for metallicities but differing in effective temperature 
determination.  The middle plots are a comparison with the values of
\citet{nordstrom04}, which were derived from Str{\"o}mgren photometry.  The lower plots compare against the values published in \citet{gray06}, derived by fitting to
low-resolution spectra.  The effective temperatures agree well with those of Valenti \& Fischer, with only two outliers that are variable stars.  There appears to be offsets
between this work and both Nordstr{\"o}m et al. and Gray et al.  The metallicity plots agree well with all three of the published work, with the lowest RMS scatter found
against the work of Valenti \& Fischer due to the similarity in the methodology.  However, a scatter of only
$\sim$0.1 was also found compared to the other methods.  The solid lines through all six plots represent the 1:1 relationship.}
\label{met_comparison_scatter}
\end{figure*}

An integral part of any abundance determination of a large sample of solar-type stars is to compare the final values with those published in other work.  
We have chosen three of the most recent large bodies of
published metallicity values in the southern hemisphere to compare against our determined abundances.  The three works we have compared against are \citet{valenti05}, 
\citet{nordstrom04} and \citet{gray06}.  The most appropriate work to compare the results found here with are those of Valenti \& Fischer as the
methodologies are similar.  They determine their final abundances by directly comparing observed high-resolution, high-S/N echelle spectra against a grid of synthetic model
spectra generated using \citet{kurucz92} model atmospheres.  This comparison is shown for both the effective
temperatures and the metallicities in the upper panels (left and right) of
Figure~\ref{met_comparison_scatter}.  The \emph{T}$_{\rm{EFF,Valenti}}$ and [Fe/H]$_{\rm{Valenti}}$ are the effective temperatures and iron abundances by Valenti \& Fischer
and the \emph{T}$_{\rm{EFF,FEROS}}$ and [Fe/H]$_{\rm{FEROS}}$ are similar but for the values found in this project.  The
solid lines represent a 1:1 relationship and it is clear that the values in both panels are in good agreement.  The effective temperatures (left panel) in Valenti \& Fischer
were derived
by a spectral fitting procedure, which is different to our photometrically derived IRFM values, however the data are in good agreement with only two outliers that are significantly
below the 1:1 relationship.  Both outliers have been found to exhibit some level of variability by Hipparcos and hence the photometric method can produce spurious results.  The RMS scatter around 
a linear fit to this data is $\pm$132K, which reduces to $\pm$80K once the two variable outliers 
have been removed.  The Student-T test reveals a 85\% probability that both these have similar means, increasing to 91\% when the outliers have been removed.  The metallicity 
comparison (right panel) shows a tight relationship between that data.  This is confirmed by the scatter of only $\pm$0.09, reducing to $\pm$0.05 once the two outliers have been 
removed.  The T-test revealed a 96\% probability that both distributions have the same means, reducing to 95\% when the outliers have been removed.  However, if we consider 
all objects with metallicities $\ge$0.1dex from both Valenti \& Fischer and here, then the Valenti \& Fischer values are found to be more metal-rich by +0.025dex.

The middle and lower panels in Figure~\ref{met_comparison_scatter} show comparisons against the two other southern samples, however the methodologies
were different to that employed here.  The middle panel shows the comparison with the values published in \citet{nordstrom04}
([M/H]$_{\rm{Nordstrom}}$).  Nordstr{\"o}m et al.
determined the metallicity of $\sim$14000 stars using relationships between the overall stellar metallicity and their Str{\"o}mgren colours.  Str{\"o}mgren photometry has
proven a useful technique in estimating a star's metallicity, however the calibrations tend to have larger systematic uncertainty than spectroscopically determined abundances
due to a lack of calibrators, particularly towards the reddest stars studied (e.g. \citealp{twarog02}).  Nordstr{\"o}m et al. created their own calibrations following
techniques previously employed by other authors (e.g., \citealp{schuster89}; \citealp{edvardsson93}) and they find a scatter of 0.12 for G and K-type stars, reducing to 0.1 
for
F-types.  It must be noted that the relations determined are compared to the iron abundances ([Fe/H]) of the calibrator stars since iron is an accurate tracer of the overall
stellar metallicity abundance.  From Fig.~\ref{met_comparison_scatter} (middle left) it can be seen that the scatter in effective temperature between this work and
Nordstr{\"o}m et al. is lower than both the other two works.  The measured RMS scatter is 73K, however a step change of $\pm$50K seems apparent at an effective temperature of 5250K.  
For the whole sample the T-test reveals a 82\%
probability that both these distributions have similar means.  However, for the lower and upper temperature cut the probabilities are 5\% and 28\% respectively.  These offsets mainly 
arise due to the different methodologies used to generate the effective temperatures i.e. Str{\"o}mgren versus Johnson filters.  The metallicity 
comparison (middle right) has a scatter of 0.13~dex and a T-test probability of 100\% that both are drawn from the same mean.  This level of agreement between the two data show that 
with a sufficiently large comparison dataset, the values in this work are as accurate as any measurements currently in the literature.

As mentioned in Section 4.1, the Gray et al. work is part of the
NStars Project and their methodology employed fitting models directly to a large spectral region of low-resolution spectra, whilst also fitting to the observed and synthetic
fluxes (see \citealp{gray03}).  The synthetic spectra were generated using \citet{kurucz93} model atmospheres and the spectral synthesis program \sc spectrum \rm.  A
Chi-squared minimisation technique was used, in conjunction with a \sc simplex \rm interpolation to generate a global fit to the observed spectra.  Again Gray et al. provides
a useful test as their method made use of stellar spectra to fit models directly.  They have also compared
their overall [M/H] values to the [Fe/H] abundances measured by \citet{valenti05} and found a scatter of only 0.09~dex, however they found an offset of -0.07~dex.
From the lower right plot in Fig.~\ref{met_comparison_scatter} it can again be seen that the derived metallicity values are in good agreement with those of Gray et al.
([M/H]$_{\rm{Gray}}$).  The scatter is found to be 0.09~dex, which compares well with the scatter found between Gray et al. and Valenti \&
Fischer.  There may be an offset of -0.01dex between these samples, however since this is below the general scatter in the data, this is deemed negligible.  This is confirmed by the 
T-test which reveals a probability of 81\%, indicating that the offset is not significant.  The lower left plot in Fig.~\ref{met_comparison_scatter} shows the comparison with the 
Gray et al. effective temperatures and a good correlation is found between the two datasets.  The scatter is found to be $\pm$90K and no significant offset was found, giving rise to 
a probability of 93\% that both datasets have the same means.  Since Gray et al. had shown their values to be in agreement with those of the IRFM, a good correlation was to be 
expected.  By comparing FEROS values to three
independent studies that employed three different methodologies it is clear that these values provide a robust assessment of the effective temperatures and iron abundances of
all stars in this sample.

\section{Results \& Discussion}

\subsection{Activity Analysis}

\begin{figure}
\vspace{3.1cm}
\hspace{-4.0cm}
\includegraphics{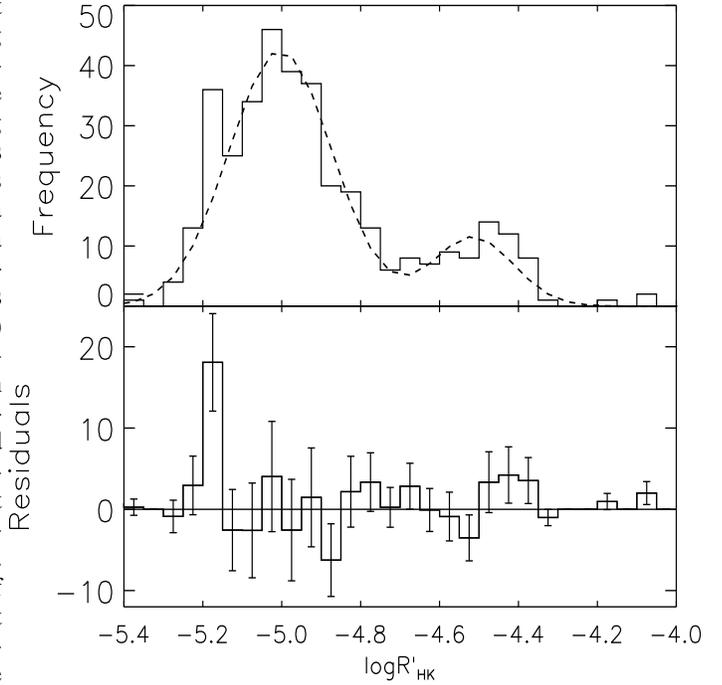}
\vspace{6.2cm}
\caption[log\emph{R}$'_{\rm{HK}}$ distribution of this sample]{The upper plot shows the log\emph{R}$'_{\rm{HK}}$ distribution of all bright stars
in the southern hemisphere in this study and the inactive and active peaks are
confirmed.  The overall distribution is in agreement with the bimodal fits from \citet{henry} and \citet{gray06} in the southern hemisphere,
however a third component may exist.  The best-fit to the data employs a bimodal distribution and this is overplotted by the dashed line.  The means of the Gaussians are
-4.52 (active) and -5.00 (inactive) with standard deviations of 0.13 and 0.10 respectively.   The lower plot
shows the residuals of the fit along with their associated error bars.  Most of the bins are in good agreement with the fit, with only one bin over
2$\sigma$ away from the fit.  This bin contains mostly stars evolving off the main sequence.
}
\label{rhk_dis}
\end{figure}

The distribution of chromospheric activity for solar-type stars has been shown to follow a bimodal distribution with both an active and inactive peak (e.g. \citealp{duncan};
\citealp{henry}; \citealp{gray06}; J06).  The upper panel in Figure~\ref{rhk_dis} shows the distribution of stellar log\emph{R}$'_{\rm{HK}}$ in this study and
confirms the bimodal peaks found before.  The plot has been binned up with bin widths of 0.05 in log\emph{R}$'_{\rm{HK}}$ and both peaks are easily identifiable.  A bimodal
gaussian has been fit to the data and is modelled by the dashed line in the plot.  The peaks of this distribution are centered at
-4.52 (active) and -5.00 (inactive) and have standard deviations of 0.13 and 0.10 respectively.  There also are a few stars out in the tails of
the distribution.  HD27442 and HD147873 are located out in the inactive wing and have log\emph{R}$'_{\rm{HK}}$ indices of -5.39 and -5.42 respectively.

The lower plot in Figure~\ref{rhk_dis} shows the residuals of the double Gaussian fit to the distribution.  The stars located out in the active tail of
the fit are single objects and therefore the residual errors bars will place them in agreement with the fit as the uncertainty is approximated using Poisson root-N,
statistics.
From these statistics alone it is hard to test whether or not these stars are undergoing an active or inactive phase of their evolution.  Almost all the bins agree well with
the fit within the uncertainties shown except for one which is over 2$\sigma$ away from the fit.  It will be shown later in the chapter that this bin, which is centered at
-5.175, contains mostly evolved stars.  It is interesting to note that this result was also
found by \cite{gray03} at around the same inactive bin of their distribution.  They concluded that this may partly be caused by a large number of Maunder Minimum stars.
However, \citet{wright04b} found that most stars announced as in Maunder Minimum phases were in fact stars evolving off the main sequence and hence the activity
methodology used to measure their indices are not entirely adequate for comparison with main sequence stars.  A trimodal fit was also applied to the data to test if the fit
was significantly improved.  The $\chi$$^{2}$ for the bimodal distribution was 30 with 23 degrees of freedom, compared with the value of 21 with 20 degrees of freedom for the
trimodal distribution.  Therefore the reduced $\chi$$^{2}$ values are 1.29 and 1.06 respectively.  Thus, it is not clear that a significant improvement is made by assuming a 
third peak exists in the data.

\subsection{Metallicity Analysis}

\begin{figure}
\vspace{1cm}
\hspace{-4.0cm}
\includegraphics{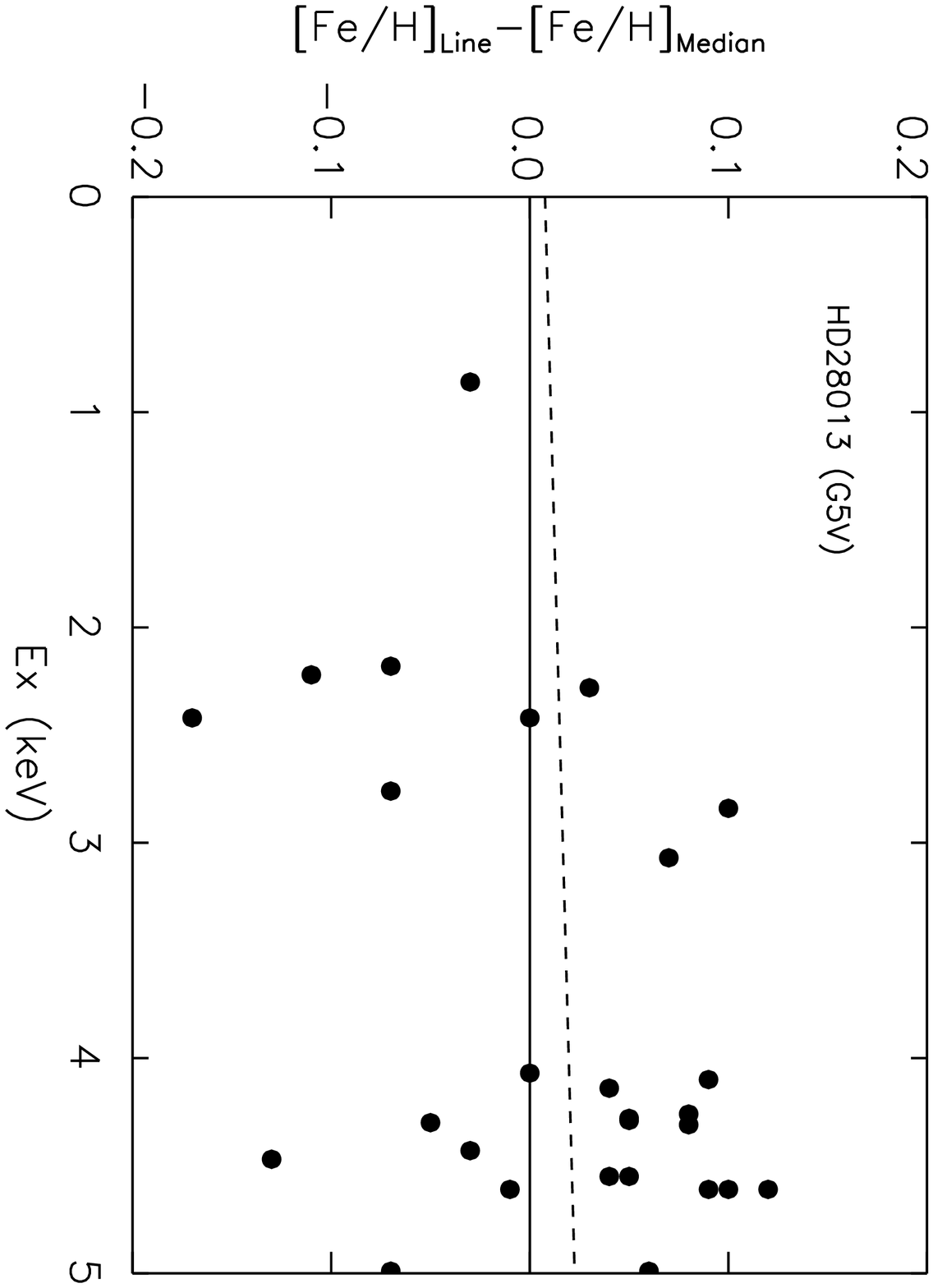}
\vspace{7.5cm}
\includegraphics{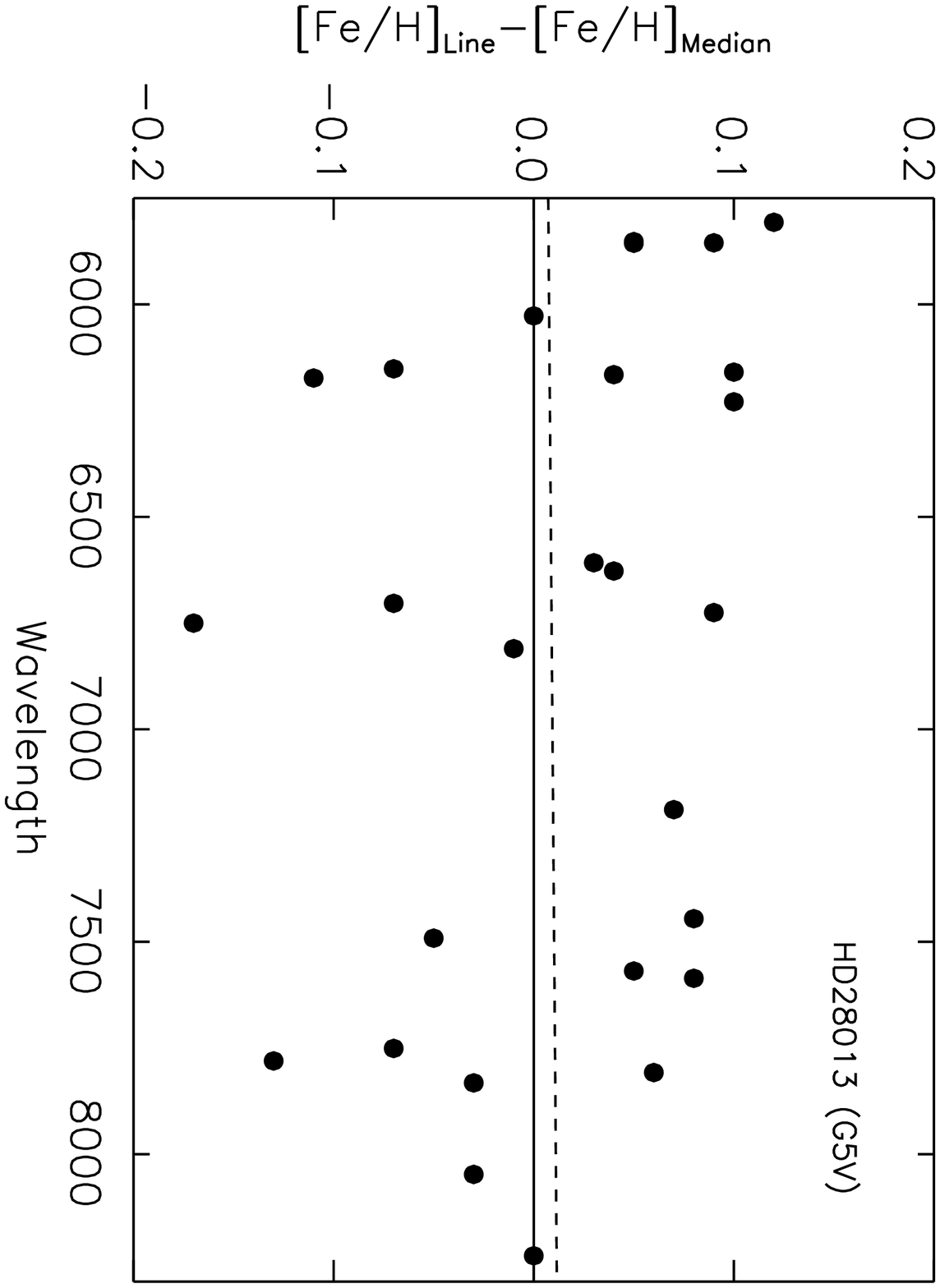}
\vspace{3.1cm}
\caption{A check of the dependency of the final abundance measurements on both absorption line excitation energy (upper panel) and wavelength (lower panel) for the star HD28013.  
These plots are typical of the whole sample.  This shows that the final abundance values are not significantly dependent on both excitation energy or wavelength.  The dashed lines 
represent best fit linear trends to the data.  The solid line marks the zero points.}
\label{ex_wav_check}
\end{figure}

To assess the validity of the metallicity abundances for each of the target stars a check of line excitation energy and wavelength dependency is performed.  The final abundance 
for each 
line is computed following the same procedure as above.  This abundance is then subtracted from each star's total abundance and plotted against both excitation 
energy and 
wavelength for the given line.  Fig.~\ref{ex_wav_check} shows both the excitation energy (upper panel) and wavelength (lower panel) abundances for each of the lines chosen here in 
the solar analogue (Spectral Type G5V; [Fe/H]=0.01) HD28013.  The majority 
of data points are spread within $\pm$0.1~dex from the final abundance value in both plots, which is represented by the solid line, and it is these clustered points that give 
rise to the measured abundance value.  A linear least-squares fit was applied to the data and these are represented by the dashed lines in both plots.  It is clear that the gradient 
of these lines are very small, indicating that there is very little dependence of these abundance measurements on excitation energy or wavelength.  
These plots are typical for the whole sample of objects and indicate that the computed [Fe/H] abundances are reliable and not a function of the selected lines. 

\begin{figure}
\vspace{4.5cm}
\hspace{-4.0cm}
\includegraphics{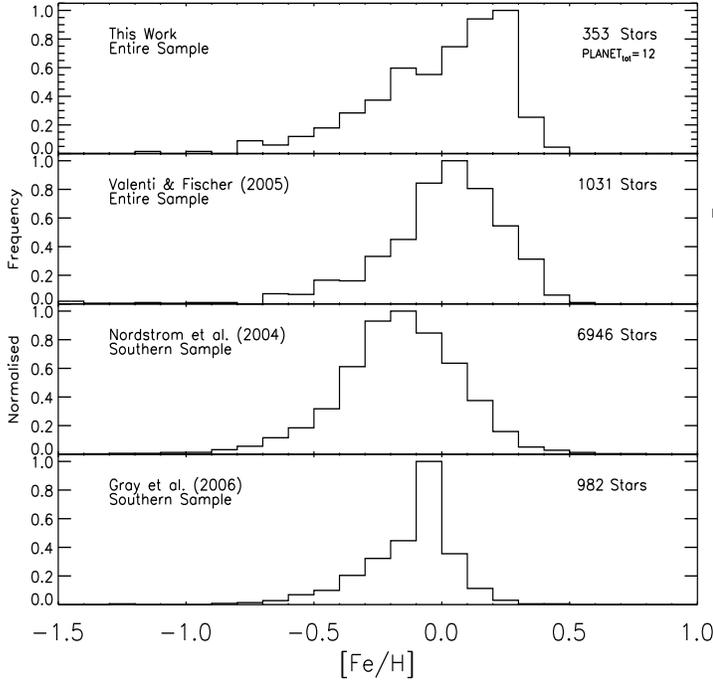}
\vspace{5.2cm}
\caption[Metallicity distributions from this work, Valenti \& Fischer (2005), Nordstr{\"o}m et al. (2004) and Gray et al. (2006)]{Four plots showing the distributions of
metallicities taken from this work (upper), \citet{valenti05} (upper middle), \citet{nordstrom04} (lower middle) and
\citet{gray06} (lower).  Since all four distributions have significantly different sample sizes all were normalised to the peak of the distribution, with the sample sizes
indicated on the plots.  The labels also represent the constituent samples as the upper, lower middle and lower panels represent the southern hemisphere, whereas the upper
middle panel is mostly the northern hemisphere but was included as this dataset represents the comparison sample with lowest internal errors.  The upper and upper middle 
distributions peak around
0-0.1~dex due to the metal-rich bias inherent in the sample selection, whereas both the middle and bottom panels should provide a better representation of the overall
metallicity distribution in the southern hemisphere.}
\label{met_comparison}
\end{figure}

The overall distribution of metallicities for all stars in this sample is shown in the upper panel of Figure~\ref{met_comparison}.  The peak of the distribution resides at
$\sim$0.0-0.1~dex, however it is hard to draw any inferences about the distribution of metallicities in the solar neighbourhood as there are specific biases in the data.  For
instance, a subsection of the data had Str{\"o}mgren colours measured by \citet{hauck} and using calibrations found in \citet{haywood} we were able to
obtain a first approximation of the overall metallicity for these objects.  We then selected a subset of the most metal-rich of these objects to follow-up with
spectroscopy to confirm their metal-rich status and further boost the numbers of metal-rich objects that will constitute the final target list.  However, as
explained in Section~3.3, the primary list of targets were objects with no known high-resolution metallicity measurements, to further enhance the database
of fundamental stellar parameters in the southern hemisphere.  Also shown in this figure is the total number of expected stars following the probability
of hosting a planet from \citet{fischer05}.  As mentioned above, Fischer \& Valenti have found that the probability of hosting a planet follows a power law
increasing to 25\% for stars with metallicities over +0.3~dex.  By assuming this power law for the sample we have estimated the total number of planets expected to be 
found in this metal-rich sample to be $\sim$12 and is shown in the upper right of the figure.  Note this is for all objects not currently on any large planet search programs.  
Of the 137 currently discovered planets around metal-rich stars, 15 have been found to transit their host star, giving rise to a fraction of $\sim$11\% that any detected planet 
around a metal-rich star will transit.  Therefore, we expect there to be at least one transiting planet in our metal-rich sample that can
be followed-up with high resolution spectroscopy to probe the atmospheres of exoplanets.  It must be noted that this only estimates the number of hot Jupiter-type planets 
down to a mass of 0.5M$_{\rm{J}}$ and a radial-velocity precision of 40ms$^{-1}$.  Any additional hot Saturns or Neptunes will not be covered by this estimate and therefore 
the true number of detectable planets should be larger than this.

From the upper part of Figure~\ref{met_comparison} it is shown that the distribution of stars in this \emph{biased} sample mostly followed a Gaussian distribution with an 
extended
metal-poor tail.  The other panels in Fig.~\ref{met_comparison} show three metallicity distributions from the surveys we have compared to in Section 4.2.  The upper middle
panel shows the distribution of
stars taken from \citet{valenti05}.  This sample contains stars mostly from the northern hemisphere, however as the northern and southern distributions are similar (e.g.
\citealp{gray03}; \citealp{nordstrom04}; \citealp{gray06}) and the methodology used was similar to the methodology used here, it can provide a useful comparison sample.  The
peak of this data agrees well with this work (upper panel) as it lies between 0~-~0.1~dex, slightly metal-rich.  This is largely again due to a bias in their sample
selection where they selected all stars on current large scale planet search surveys, which itself introduces a number of metal-rich stars as most surveys now have metal-rich
programs.  Also the extended metal-poor tail is apparent in this data.  However, when compared with the lower middle
(\citealp{nordstrom04}) and lower (\citealp{gray06}) panels the distributions differ in where the peak lies.  Both these samples are taken solely from the southern hemisphere
(dec~$\le$0$^{\rm{o}}$) and both have peaks below solar metallicity, with the largest sample from Nordst{\"o}m et al. peaking between -0.1 and -0.2.  These samples are both
unbiased samples as their selection criteria is volume limited out to 40pc.  Nordstr{\"o}m et al. selected a magnitude limited sample of stars with Str{\"o}mgren photometry
that is complete for all F and G stars out to 40pc and is kinematically unbiased, whereas the Gray et al sample is as yet incomplete but also focused on all stars earlier
than M0 in the Hipparcos catalogue.  Both these studies should better represent the overall distribution of stars in the galactic disk.  It is clear that the Nordstr{\"o}m et
al. distribution follows a Gaussian profile with only a very small indication of a metal-poor tail, yet the Gray et al. profile shows a clear extended tail.  The extended
tail in the distributions arise mostly from stars from both the thick disk and the galactic halo and possibly from active K stars (\citealp{gray06}).  

\begin{figure}
\vspace{4.5cm}
\hspace{-4.0cm}
\includegraphics{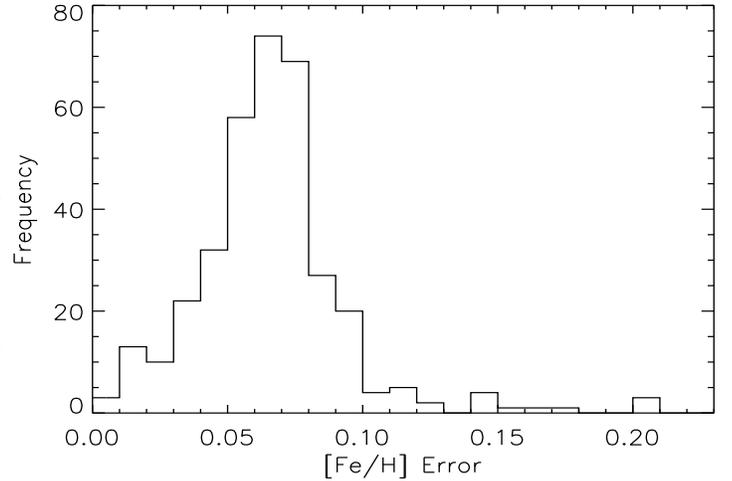}
\vspace{1.7cm}
\caption[The distribution of $\chi$$^{2}$ errors on the metallicity measurements]{The distribution of 1-$\sigma$ uncertainties from the chi-squared fitting for all
stars in this project.  The distribution peaks at $\pm$0.065~dex, with the majority of values spread between 0.03-0.10.  There are also a number of stars with larger
uncertainties, which were mostly the stars that were observed through cloud cover and have lower S/N.}
\label{met_err_dis}
\end{figure}

Figure~\ref{met_err_dis} shows the distribution of all uncertainties for the metallicity fits to these stars in this sample.  The uncertainties appear to follow a
normal distribution peaking at $\pm$0.065~dex, which agrees well with the scatter found in the metallicities when compared to other works in the literature.  However, an
extended tail is found at larger uncertainty values.  These are mostly comprised of low S/N stars that
were observed through cloud cover.  The low S/N means there are less photons making up each Fe line and the line profiles are not well behaved and are harder to model.

\subsection{Evolutionary Phase}

\begin{figure}
\vspace{6.5cm}
\hspace{-4.0cm}
\includegraphics{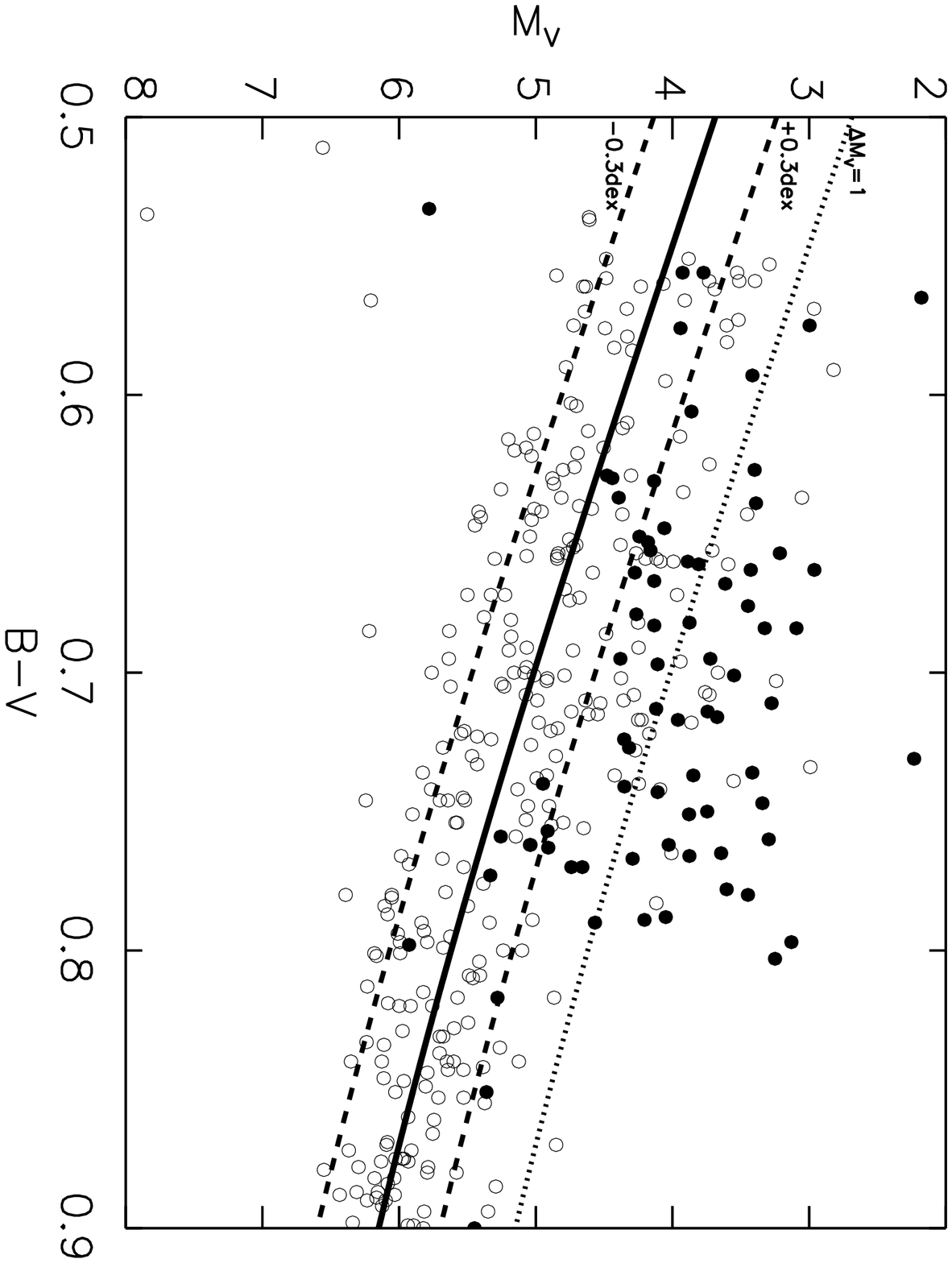}
\includegraphics{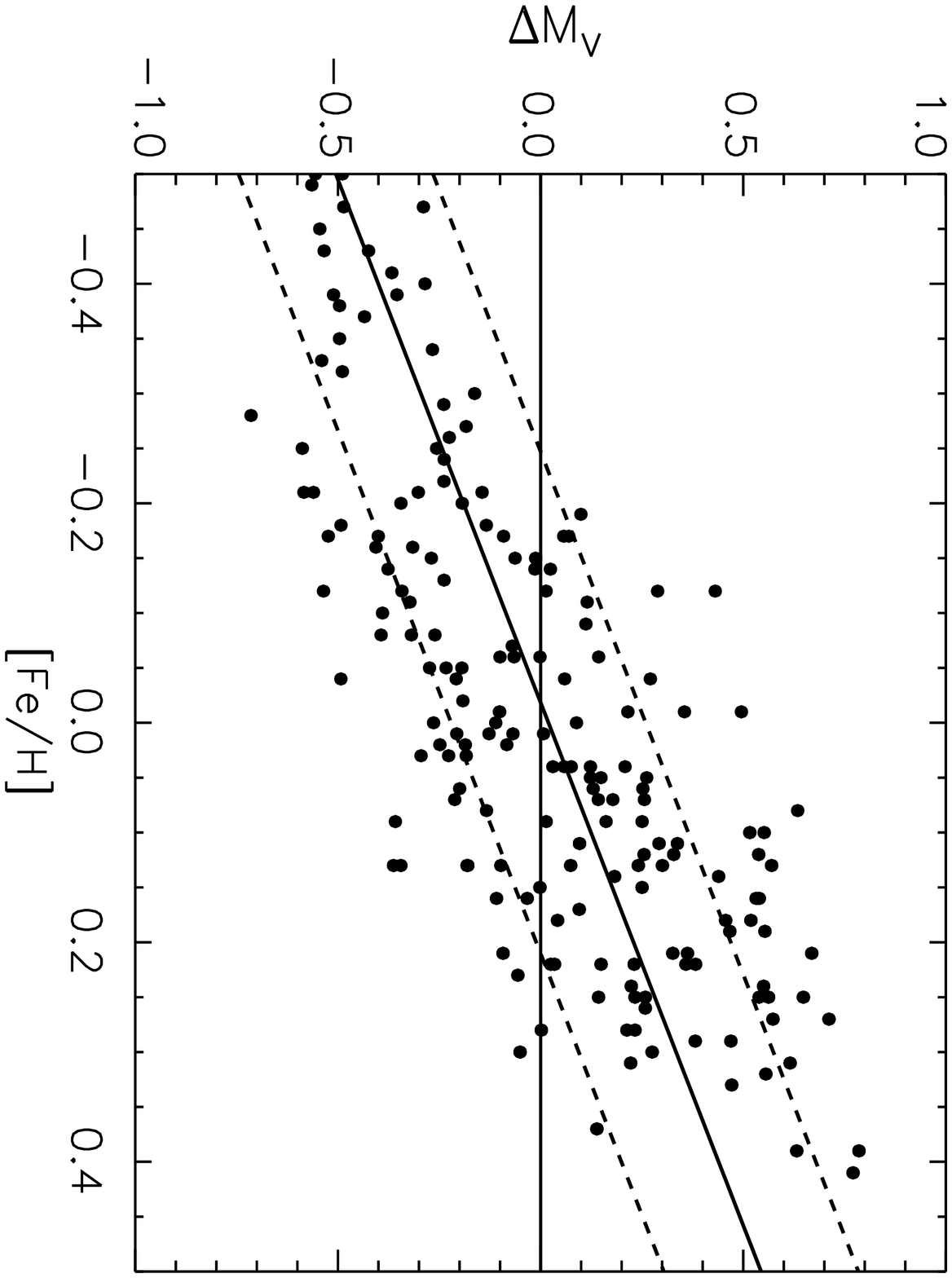}
\vspace{4.5cm}
\caption[Evolutionary status for all stars in this study]{The top plot shows a Hipparcos colour-magnitude diagram for all the stars in this survey except for the calibration
stars above a $B-V$ of 0.9.  The open circles represent all the target
stars in this colour space.  The solid line is the empirical main sequence for Hipparcos stars determined in \citet{wright04b}, with the dashed lines representing the spread
in the main sequence due to metallicity ($\pm$0.3dex).  The dotted line represents a difference of 1~magnitude above the main sequence and this isolates evolved stars.  It is
a clear a number of these stars are evolving off the main sequence.  There are also a number of a stars far below the main sequence that comprise thick-disk and halo
stars.  The filled circles represent possible Maunder Minimum stars in this study (upper panel).  The lower plot shows the distance from the fitted main sequence against each
stars
metallicity within a subset of the original sample.  This represents all stars where the surface gravity was greater than 4.0 (log(\emph{g})).  The solid line represents the
best-fit straight line with the dashed lines representing the 1$\sigma$ uncertainty boundaries.  This fit can be used to correct each stars position on the main sequence
for the effects of metallicity.}
\label{activity_hr}
\end{figure}

To determine if any of the stars in this project are in a Maunder Minimum phase of their evolution we have tried to ascertain their current evolutionary status.
\citet{baliunas90} determined that 30\% of the MW stars were in a Maunder minimum phase, this was based purely on their low activity status.  This estimate was revised to a
lower percentage of 10-15\% by \citet{saar92}, which has since been complimented by other studies (\citealp{henry}; \citealp{saar98}; \citealp{gray03}).  However, \citet{wright04b}
has shown that these stars are all either evolved, so the application of the activity index does not apply, or are metal-poor, which alters the CaHK line
wings for such stars so their indices are unreliable.  Note that activity indices are calibrated for stars thought to be on the main sequence and hence the further from the 
main
sequence a star evolves the more unreliable the index becomes.  Indeed it may be necessary to re-determine the activity indices for more evolved stars.  What is clear is that
the Sun went through a similar activity minimum in the 17$^{\rm{th}}$ Century and it has been estimated that the solar \emph{S}$_{\rm{MW}}$ index
decreased to a mean value of around 0.145 (\citealp{baliunas90}) (currently the mean is \emph{S}$_{\odot}$=0.165).  By applying this determination to the analysis of
brightness changes due to activity, \cite{zhang94} have determined that the Sun was 0.2\%-0.6\% dimmer during the Maunder Minimum.  Stars in this 
study may be in a similar evolutionary phase.  Figure~\ref{activity_hr} shows the distribution of
stars in this survey on a colour-magnitude diagram.  The $B-V$ colours and absolute magnitudes were extracted from the Hipparcos catalogue (\citealp{perryman}).  The main
sequence is clear in this plot, however to obtain the true main sequence we used the model of \citet{wright04b}.  This model was determined using a 9$^{\rm{th}}$-order
polynomial fit to the Hipparcos main sequence, described by:

\begin{equation}
\label{eq:main_sequence}
M_{V,MS}(B-V)=\sum a_{i}(B-V)^{i}
\end{equation}

Here, $M_{V,MS}(B-V)$ is the absolute magnitude of the main sequence as a function of colour, a$_{i}$ are the coefficients describing the polynomial, which are the
following: 0.909, 6.258, -23.022, 125.5537, -321.1996, 485.5234, -452.3198, 249.6461, -73.57645 and 8.8240.  As expected the majority of stars are clustered around the main
sequence fit, with a few outliers above and below the fit.  The majority of stars far below the fit are extremely metal-poor objects showing that metallicity can
significantly alter a stars position on the HR-diagram.  Conversely, high metal abundance will place stars above the main sequence and contribute to the spread of values
around the fitted main sequence.  The dashed lines represent the spread across the main sequence for a metallicity of one-third the solar metallicity (lower) and three times
solar (upper) in a similar fashion to the analysis of Wright.  Stars that are significantly higher than the main sequence are evolving off the main sequence and moving across 
to the red giant branch.  Wright has shown
that the majority of stars announced as being in a Maunder minima are actually evolved stars and hence the indices are not applicable to this conclusion.  There are 79 stars
in this study that are extremely inactive (log\emph{R}$'_{\rm{HK}}\le$-5.10).  Other authors have claimed that stars with such properties are in a Maunder minimum phase 
(e.g. \citealp{henry}; \citealp{saar98}; \citealp{gray03}).  These stars are highlighted in Fig.~\ref{activity_hr} by filled circles.  It is clear
that the majority are located high above the main sequence and are evolving onto the giant branch, with two below and the rest close to the main sequence.  To determine
how many of these stars are clearly on the main sequence we use the definition of the height above the main sequence in \citet{wright04b}.

\begin{equation}
\label{eq:delta_mv}
(\Delta M_{V})_{MS,B-V}=M_{V,MS}(B-V)-M_{V}
\end{equation}

($\Delta M_{V}$)$_{MS,B-V}$ is the difference between a stars absolute magnitude and the absolute magnitude of the main sequence fit (for that $B-V$).  Eq.~\ref{eq:delta_mv} 
is used to determine the
height above the main sequence, giving an indication if a star is evolving off the main sequence or not.  One must still be careful when working with solitary
stars around the main sequence as their position is metallicity sensitive with stars of low metallicity lying below the main sequence and stars with high metallicities lying
above the main sequence.  Wright estimates that stars with a metallicity of +0.3~dex will lie $\sim$0.45 magnitudes above the main sequence, whereas stars with a metallicity
of -0.3~dex will lie $\sim$0.45 magnitudes below the main sequence.  These regions are shown as dashed lines in Fig.~\ref{activity_hr} and the majority of stars in this study
are located between these two bounds, both inactive and active stars.  Also shown on the plot is the region that marks ($\Delta M_{V}$)$_{MS,B-V}$=1 (dotted line) which we 
use as a
boundary for evolutionary status.  Stars above this line we expect to be in an evolved state of their evolution regardless of their metallicity.  Since we have measured the 
metallicities, a correction can be made to remove the metallicity component in each star's main sequence position.

The lower plot in Figure~\ref{activity_hr} shows the distribution of stellar metallicities against distance from the main sequence.  A clear trend is seen in the plot with
the metal-poor objects generally found below the main sequence and metal-rich objects located above.  This sequence was cleaned by focusing on stars that show no significant
evidence for evolution into giant status.  As mentioned above, all stars with a ($\Delta M_{V}$)$_{MS,B-V}~>$1 must be evolving off the main sequence and are removed from 
this sample.
To ensure this is the case and to detect any evolved objects located ($\Delta M_{V}$)$_{MS,B-V}~<$1, an estimate of the surface gravity was made following the technique described in 
Section~3.3.1.  Using surface gravities we were able to remove all stars
with a log(\emph{g}) of 3.5 or 4.0, which with grid steps of 0.5, should effectively remove all stars below 4.25.  This is a conservative estimate but should
select against any stars evolving off the main sequence and leave behind only true main sequence stars.  A first fit was then made to this subset of stars to estimate where
the main sequence lies and the 1$\sigma$ scatter was measured and used to perform a secondary cut, which removed all objects outside the first scatter region.  This cut was
used to better constrain the position of the main sequence by focusing on the region that is most populated.  The thick solid line Fig.~\ref{activity_hr} (lower) shows the
best-fit linear trend to this main sequence across all metallicities, after all cuts have been performed, accompanied by the 1$\sigma$ uncertainty regions (dashed lines).
The trend is evident and it should be noted that the fit intersects the point of zero distance from the main sequence very close to the solar metallicity value
of zero without any necessary fixing or adjustment.  The relationship is described by:

\begin{equation}
\label{eq:delta_mv_met}
(\Delta M_{V})_{[Fe/H]}=1.050([\rm{Fe/H}])+0.019
\end{equation}

The uncertainties on the fitted coefficients in Equation.~\ref{eq:delta_mv_met} are 0.018 and 0.071 respectively, and the 1$\sigma$ uncertainty shown by the dashed lines in
the figure is $\pm$0.240.  Since the
majority of stars reside within the scatter region we are confident this region is a good proxy for the main sequence in this sample.  Simply by subtracting off 
($\Delta$ M$_{V}$)$_{[Fe/H]}$ from Eq$^{n}$.~\ref{eq:delta_mv_met} from the initially determined ($\Delta$M$_{V}$)$_{MS,B-V}$ in
Eq$^{n}$.~\ref{eq:delta_mv} we can remove the effects of metallicity on the spread in distance from the main sequence.  This repositions all stars onto the main
sequence irrespective of their metallicities and since all evolved objects have been removed a clean sample of possible Maunder minimum candidates can be made, based on
their inactivity and distance from the main sequence.  From the stars that are left we have selected those that are within the 1$\sigma$ boundaries of this fitted main
sequence.  From this selection six stars remain and should represent both main sequence stars and very inactive stars, all of these are listed in Table.~\ref{tab:maunder}.
A check of any affects due to line strength changes between metal-poor and metal-rich objects was made to test if any lines within the bandpass regions were biasing
the metal-rich stars towards lower activities.  One absorption line was found inside the H bandpass region and none inside the K bandpass region.  We interpolated this line
out and
recomputed the activity index and found that there were no real changes in the final log\emph{R}$'_{\rm{HK}}$
activities.  This result was similar for the continuum bandpass regions and therefore any line strength alterations are minimal compared with the reduction uncertainties.
However, the metallicity does affect the intrinsic colour of stars and should affect the derived log\emph{R}$'_{\rm{HK}}$ value as this is colour dependent.  It
is also important to note that since only one measurement
has been made for these objects in this study, it is unsure whether or not they will remain stable over a number of years or will undergo any cyclical variation.  If we 
assume a Gaussian distribution to describe the cyclical activity variations and take the stars with the potential to scatter above the -5.10 activity boundary (i.e. stars 
within the range -5.25 to -5.10 which represents a maximum cycle of 0.15 in log\emph{R}$'_{\rm{HK}}$ and is consistent with the Sun's cycle) then four of the six stars are 
expected to be normal stars undergoing their usual activity cycle at the 1$\sigma$ level of confidence.  Of the entire sample of 353 stars, this leads to a fraction of 
$\sim$2$\pm$1\% and hence the Sun should spend only a few percent of its main sequence lifetime in a Maunder minimum like state.  It should be noted that the sample is 
biased towards the most metal-rich objects in the solar neighbourhood.  This selection will skew the final estimate of Maunder minimum stars as it should be biased towards 
younger more active objects.

\begin{table}
\center
\caption[Maunder minimum stars]{Six possible Maunder Minimum stars, along with their Hipparcos $B-V$ colours and $V$-band magnitudes, ($\Delta$ M$_{V}$)$_{MS,B-V}$ from the 
main sequence \emph{after} metallicity correction, log\emph{R}$'_{\rm{HK,FEROS}}$ indices and metallicity values.}. \label{tab:maunder}
\
\begin{tabular}{cccccc}
\\
~~~~~~~~~~~~~~~~~~~~
                                    &      &               &                        &         \\
\hline
\multicolumn{1}{c}{HD} & \multicolumn{1}{c}{B-V} & \multicolumn{1}{c}{V} & \multicolumn{1}{c}{$\Delta$M$_{V}$} & \multicolumn{1}{c}{log\emph{R}$'_{\rm{HK,FEROS}}$} & \multicolumn{1}{c}{[Fe/H]} \\ \hline
                \underline{Data}                &       &         &   &                        &      \\

HD8446 & 0.661 &  8.16 &  -0.15 &  -5.17 &  -0.35 \\
HD9175 & 0.656 &  8.01 &  -0.24 &  -5.12 &  -0.12 \\
HD22104 & 0.679 &  8.32 &   0.07 &  -5.15 &   0.15 \\
HD31827 & 0.770 &  8.26 &   0.07 &  -5.11 &  -0.34 \\
HD84501 & 0.663 &  8.26 &  -0.03 &  -5.12 &  -0.22 \\
HD90722 & 0.724 &  7.88 &  -0.15 &  -5.10 &  -0.20 \\

\hline
\end{tabular}
\medskip
\end{table}


\subsection{Benchmark Target List}

The main aim of this study was to determine accurate chromospheric activity and metallicity values for a number of close-by, bright stars in the southern hemisphere.  The 
metal-rich subset of these will provide excellent candidates for planet search studies searching for hot giant planets.  Figure~\ref{met_comparison} (upper) shows the 
distribution of metallicities in this sample.  About 30\% of stars have metallicities greater than 0.1~dex, of which 15\% have metallicities greater than 0.2~dex and it
is these stars that represent the best stars for radial-velocity follow-up.  First epoch radial-velocities have already been acquired for a number of these stars in ESO 
Period 76 with the HARPS instrument.

\begin{figure}
\vspace{3.0cm}
\hspace{-4.0cm}
\includegraphics{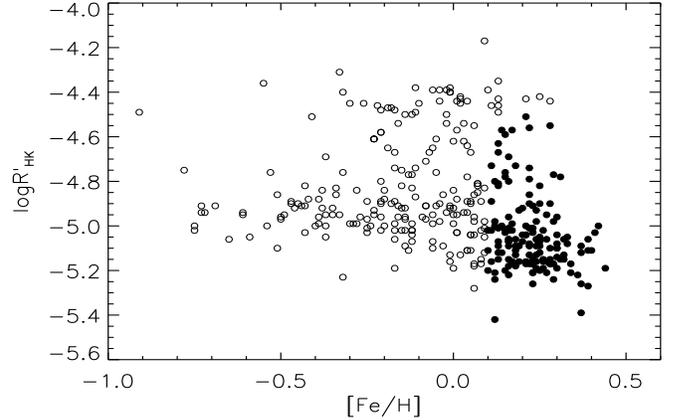}
\vspace{2.8cm}
\caption[Metallicity v's chromospheric activity of these stars]{Iron abundance as a function of chromospheric activity for the stars in this sample.  The filled circles
represent all stars that will compose our metal-rich target list.  While most stars above +0.1~dex are inactive, a few have activities above -4.5.}
\label{rhk_met}
\end{figure}

The chromospheric activities also play an important role in the nature of the objects that can be detected.  Figure~\ref{rhk_met} shows the distribution of activities
across all metallicity space.  Only a few of the metal-rich candidates are active and will be removed from the final candidate list, whereas most of the stars above +0.1dex
in [Fe/H] have log\emph{R}$'_{\rm{HK}}$ values less than -4.5.  Since
the level of radial-velocity noise (aka. \emph{jitter}) is correlated with activity (\citealp{saar98}; \citealp{santos00}; \citealp{wright05}) these inactive stars allow 
higher precision radial-velocities to be efficiently measured.  The potential to access lower amplitude signals allows access to objects of lower masses and as the mass 
function rises towards lower masses following a power law of -1.8 (\citealp{grether06}), the probability of planet detection significantly increases.  The final target list 
will be a subset of stars taken from Table.~\ref{tab:activity1} following the criteria explained above.

\section{Conclusions}

Using the high-resolution echelle data from the FEROS instrument we extracted chromospheric activities and metallicities
for 353 bright stars in the Southern hemisphere.  We have shown the activity distribution can be
described well by a bimodal function which agrees well with previous work in both the northern and southern hemispheres for such stars.  The sample is dominated by G stars
(\emph{T}$_{\rm{EFF}}$$\sim$6000K) and the
activity distribution peaks around -5.0.  The K stars have an active to inactive peak ratio close to 1, compared with
the G stars peak-to-peak ratio of 0.2.  The metallicity distribution peaks around 0.1~dex, contrary to unbiased
surveys which peak around -0.1~dex.  116 stars have metallicities greater than 0.1~dex and are not found on other planet search lists.  It is these stars that 
constitute our planned metal-rich planet search project.  The uncertainties follow a Gaussian distribution peaking at $\pm$0.085~dex with a high uncertainty tail of
objects due to stars with low S/N spectra.
The values are in agreement with similar works in the literature, such as \citet{valenti05}, with an RMS
scatter of $\sim$0.05~dex.

\begin{acknowledgements}
We would like to thank Pierre Maxted for his suggestions regarding various aspects of the paper, along with the anonymous referee for their various suggestions.  We would also like to 
acknowledge the help from Patrick Francois whilst observing with the FEROS instrument.  We also acknowledge the use of the Vizier and Simbad databases.
\end{acknowledgements}

\begin{table*}
\center
\caption[FEROS calibration activity table]{Derived activity and metallicity values for all stars used to calibrate onto the Mt.~Wilson system of measurements.
HD Number refers to the Henry Draper catalogue identifier.  Two stars have no HD catalogue reference and are catalogued by their Hipparcos identifier (HIP).  Johnson B-V, V 
taken from the Hipparcos catalogue.  \emph{S}$_{\rm{FEROS,a}}$ is \emph{before} calibration onto the Mt.~Wilson and \emph{S}$_{\rm{FEROS,b}}$ is \emph{after} calibration.   
\emph{S}$_{\rm{MW}}$ is the Mt.~Wilson index used to perform this calibration, taken from column 5 of table 1 and
columns 4-7 of table 3 from \citet{duncan}.  log\emph{R}$'_{\rm{HK,MW}}$ are derived from \citet{duncan}.  Note that the photon-counting uncertainty in the derived activities 
are indicative
only; based on measurements of $\tau$~Ceti a random uncertainty of $\ga$~3\% must be taken into account in all instances.  [Fe/H] is used to denote iron abundances (herein 
referred to as metallicity).}.
\label{tab:calibrators1}
\
%

\begin{tabular}{ccccccccc}
\hline
\multicolumn{1}{c}{HD} & \multicolumn{1}{c}{B-V} & \multicolumn{1}{c}{V} & \multicolumn{1}{c}{\emph{S}$_{\rm{FEROS,a}}$}  & \multicolumn{1}{c}{\emph{S}$_{\rm{FEROS,b}}$}
& \multicolumn{1}{c}{\emph{S}$_{\rm{MW}}$} & \multicolumn{1}{c}{log\emph{R}$'_{\rm{HK,FEROS}}$} & \multicolumn{1}{c}{log\emph{R}$'_{\rm{HK,MW}}$} &
\multicolumn{1}{c}{[Fe/H]} \\ \hline

                \underline{Calibrators} &  &             &       &    &    &   &          &              \\

HD1835     & 0.659     &  6.39     & 0.291   &     0.350$\pm$0.003 &   0.347     & -4.43 & -4.44     &     0.21$\pm$0.03    \\
HD3443AB     & 0.715     &  5.57   &  0.142  &     0.180$\pm$0.001 & 0.186     & -4.92 & -4.89     &    -0.01$\pm$0.05    \\
HD3795     & 0.718     &  6.14     &  0.119  &  0.153$\pm$0.001 &   0.156     & -5.06 & -5.04     &    -0.65$\pm$0.05    \\
HD9562     & 0.639     &  5.75     &  0.106  &  0.138$\pm$0.001 &   0.144     & -5.16 & -5.11     &     0.19$\pm$0.14    \\
HD10700     & 0.727     &  3.49     & 0.140  &  0.178$\pm$0.001 &  0.173     & -4.93 & -4.95     &    -0.38$\pm$0.04    \\
HD10700     & 0.727     &  3.49     & 0.142  &  0.180$\pm$0.001 &  0.173     & -4.92 & -4.95     &    -0.35$\pm$0.04    \\
HD22484     & 0.575     &  4.29     & 0.118  &  0.153$\pm$0.001 &  0.147     & -5.02 & -5.06     &    -0.14$\pm$0.02    \\
HD30495     & 0.632     &  5.49     & 0.234  &  0.286$\pm$0.002 &  0.292     & -4.54 & -4.54     &     0.03$\pm$0.03    \\

\hline
\end{tabular}
\medskip
\end{table*}

\begin{table*}
\center
\caption[FEROS activity table]{\emph{S$_{\rm{FEROS}}$} and log\emph{R}$'_{\rm{HK,FEROS}}$ for bright solar-type stars in the southern hemisphere.  HD Number refers to the 
Henry Draper catalogue identifier.  Johnson B-V, V and Spec Type are all taken from the Hipparcos catalogue.
\emph{S}$_{\rm{FEROS}}$ is the FEROS activity index \emph{after} calibration onto the Mt. Wilson
system of measurements.  log\emph{R}$'_{\rm{HK,FEROS}}$ are final activity values generated following the \citet{noyes} methodology.  [Fe/H] is the iron abundance.}
\label{tab:activity1}
\

\begin{tabular}{cccccc}
\hline
\multicolumn{1}{c}{HD/HIP}& \multicolumn{1}{c}{B-V}& \multicolumn{1}{c}{V}  & \multicolumn{1}{c}{\emph{S}$_{\rm{FEROS}}$}
& \multicolumn{1}{c}{log\emph{R}$'_{\rm{HK,FEROS}}$} & \multicolumn{1}{c}{[Fe/H]} \\ \hline
                                           &       &   &                        &   &   \\
                \underline{Measurements}           &     &                                     &   &   &  \\

HD2595     & 0.840     &  9.34     &   0.174$\pm$0.001     & -5.00     &    -0.40$\pm$0.07    \\
HD3569     & 0.846     &  9.21     &   0.176$\pm$0.001     & -4.99     &    -0.29$\pm$0.08    \\
HD5185     & 0.722     &  8.92     &   0.239$\pm$0.002     & -4.71     &    -0.08$\pm$0.07    \\
HD6348     & 0.801     &  9.15     &   0.177$\pm$0.001     & -4.97     &    -0.50$\pm$0.06    \\
HD6558     & 0.606     &  8.20     &   0.141$\pm$0.001     & -5.12     &     0.25$\pm$0.06    \\
HD6790     & 0.561     &  9.37     &   0.168$\pm$0.001     & -4.92     &     0.20$\pm$0.03    \\
HD7112     & 0.672     &  8.80     &   0.382$\pm$0.003     & -4.39     &    -0.04$\pm$0.04    \\
HD7320     & 0.685     &  8.95     &   0.163$\pm$0.001     & -4.99     &    -0.28$\pm$0.06    \\
HD7621     & 0.762     &  9.03     &   0.142$\pm$0.001     & -5.13     &    -0.01$\pm$0.07    \\
HD7950     & 0.797     &  8.72     &   0.139$\pm$0.001     & -5.16     &     0.17$\pm$0.07    \\
HD8038     & 0.701     &  8.41     &   0.210$\pm$0.002     & -4.79     &     0.22$\pm$0.06    \\
HD8389     & 0.900     &  7.85     &   0.152$\pm$0.001     & -5.11     &     0.40$\pm$0.08    \\
HD8446     & 0.661     &  8.16     &   0.137$\pm$0.001     & -5.17     &     0.28$\pm$0.04    \\
HD8765     & 0.707     &  8.14     &   0.180$\pm$0.001     & -4.91     &     0.23$\pm$0.06    \\
HD9279     & 0.630     &  8.15     &   0.164$\pm$0.001     & -4.97     &    -0.08$\pm$0.04    \\
HD9796     & 0.780     &  8.77     &   0.179$\pm$0.001     & -4.95     &    -0.37$\pm$0.07    \\
HD10008     & 0.797     &  7.66     &  0.468$\pm$0.004     & -4.39     &   -0.02$\pm$0.06        \\
HD10188     & 0.676     &  8.94     &  0.140$\pm$0.001     & -5.14     &     0.30$\pm$0.06    \\
HD10278     & 0.695     &  9.13     &  0.279$\pm$0.002     & -4.59     &     0.15$\pm$0.07    \\
HD11352     & 0.644     &  8.61     &  0.308$\pm$0.003     & -4.50     &    -0.12$\pm$0.04    \\
HD11683     & 0.887     &  9.15     &  0.508$\pm$0.004     & -4.45     &    -0.26$\pm$0.08    \\
HD12754     & 0.566     &  8.66     &  0.206$\pm$0.002     & -4.74     &    -0.11$\pm$0.14    \\
HD12873     & 0.795     &  8.95     &  0.151$\pm$0.001     & -5.09     &    -0.14$\pm$0.07    \\
HD12889     & 0.885     &  8.55     &  0.167$\pm$0.001     & -5.05     &    -0.37$\pm$0.07    \\
HD13147     & 0.695     &  9.21     &  0.136$\pm$0.001     & -5.18     &     0.16$\pm$0.07    \\
HD13350     & 0.766     &  8.94     &  0.131$\pm$0.001     & -5.21     &     0.34$\pm$0.08    \\
HD13977     & 0.880     &  9.11     &  0.244$\pm$0.002     & -4.82     &     0.07$\pm$0.10    \\
HD14655     & 0.684     &  8.13     &  0.138$\pm$0.001     & -5.16     &    -0.01$\pm$0.05    \\
HD15191     & 0.704     &  8.65     &  0.193$\pm$0.002     & -4.85     &     0.07$\pm$0.06    \\
HD15507     & 0.670     &  8.66     &  0.178$\pm$0.001     & -4.90     &     0.22$\pm$0.06    \\
HD17194     & 0.803     &  8.71     &  0.122$\pm$0.001     & -5.28     &     0.06$\pm$0.08    \\
HD18241     & 0.672     &  8.66     &  0.346$\pm$0.003     & -4.45     &    -0.10$\pm$0.03    \\
HD18599     & 0.876     &  8.99     &  0.527$\pm$0.004     & -4.42     &    -0.01$\pm$0.12    \\
HD18599     & 0.876     &  8.99     &  0.544$\pm$0.005     & -4.40     &    -0.01$\pm$0.13    \\
HD18708     & 0.651     &  8.42     &   0.136$\pm$0.001     & -5.18     &     0.18$\pm$0.05    \\
HD18754     & 0.736     &  8.44     &   0.131$\pm$0.001     & -5.22     &     0.16$\pm$0.06    \\
HD19493     & 0.695     &  8.14     &   0.140$\pm$0.001     & -5.15     &     0.23$\pm$0.06    \\
HD19773     & 0.551     &  8.06     &   0.177$\pm$0.001     & -4.86     &     0.22$\pm$0.03    \\
HD20605     & 0.766     &  9.22     &   0.176$\pm$0.001     & -4.95     &    -0.49$\pm$0.06    \\
HD22104     & 0.679     &  8.32     &   0.139$\pm$0.001     & -5.15     &     0.31$\pm$0.05    \\
HD22177     & 0.717     &  7.78     &   0.152$\pm$0.001     & -5.06     &     0.16$\pm$0.06    \\
HD22282     & 0.789     &  8.54     &   0.164$\pm$0.001     & -5.02     &     0.16$\pm$0.07    \\
HD22472     & 0.784     &  8.85     &   0.229$\pm$0.002     & -4.78     &     0.04$\pm$0.09    \\
HD23065     & 0.751     &  8.25     &   0.156$\pm$0.001     & -5.05     &    -0.59$\pm$0.04    \\
HD23398     & 0.788     &  8.43     &   0.133$\pm$0.001     & -5.19     &     0.44$\pm$0.07    \\
HD24091     & 0.817     &  8.89     &   0.485$\pm$0.004     & -4.39     &    -0.06$\pm$0.03    \\
HD25015     & 0.899     &  8.87     &   0.537$\pm$0.005     & -4.44     &     0.04$\pm$0.07    \\
HD25061     & 0.843     &  9.27     &   0.238$\pm$0.002     & -4.80     &     0.12$\pm$0.08    \\
HD26071     & 0.742     &  8.89     &   0.157$\pm$0.001     & -5.04     &     0.21$\pm$0.06    \\
HD26965     & 0.820     &  4.43     &   0.181$\pm$0.001     & -4.96     &    -0.27$\pm$0.09    \\
HD27905     & 0.627     &  7.82     &   0.174$\pm$0.001     & -4.91     &    -0.08$\pm$0.06    \\
HD28013     & 0.701     &  8.38     &   0.186$\pm$0.001     & -4.88     &     0.01$\pm$0.05    \\
HD28255     & 0.659     &  6.28     &   0.171$\pm$0.001     & -4.94     &     0.08$\pm$0.04    \\
HD28255     & 0.659     &  6.28     &   0.179$\pm$0.001     & -4.89     &     0.05$\pm$0.04    \\
HD29231     & 0.776     &  7.61     &   0.246$\pm$0.002     & -4.73     &     0.11$\pm$0.07    \\
HD29813     & 0.621     &  7.74     &   0.173$\pm$0.001     & -4.91     &    -0.05$\pm$0.05    \\
HD30501     & 0.875     &  7.58     &   0.334$\pm$0.003     & -4.65     &    -0.06$\pm$0.08    \\
HD31827     & 0.770     &  8.26     &   0.147$\pm$0.001     & -5.11     &     0.39$\pm$0.07    \\
HD33449     & 0.681     &  8.49     &   0.180$\pm$0.001     & -4.90     &    -0.47$\pm$0.11    \\
HD33725     & 0.799     &  8.04     &   0.220$\pm$0.002     & -4.82     &    -0.15$\pm$0.07    \\
HD34688     & 0.834     &  9.08     &   0.203$\pm$0.002     & -4.89     &    -0.21$\pm$0.07    \\
HD34962     & 0.733     &  8.27     &   0.383$\pm$0.003     & -4.44     &    -0.04$\pm$0.05    \\
HD36169     & 0.702     &  8.32     &   0.170$\pm$0.001     & -4.96     &    -0.09$\pm$0.05    \\

\hline
\end{tabular}
\medskip
\end{table*}

\begin{table*}
\center
\begin{tabular}{cccccc}
\hline

HD37634     & 0.647     &  8.91     &   0.186$\pm$0.001     & -4.86     &    -0.51$\pm$0.03    \\
HD38459     & 0.861     &  8.52     &   0.471$\pm$0.004     & -4.46     &     0.13$\pm$0.06    \\
HD38467     & 0.672     &  8.27     &   0.147$\pm$0.001     & -5.09     &     0.22$\pm$0.05    \\
HD39833     & 0.629     &  7.65     &   0.203$\pm$0.002     & -4.78     &     0.15$\pm$0.04    \\
HD39997     & 0.559     &  8.48     &   0.161$\pm$0.001     & -4.96     &     0.06$\pm$0.03    \\
HD40293     & 0.720     &  9.10     &   0.206$\pm$0.002     & -4.81     &     0.13$\pm$0.06    \\
HD42538     & 0.593     &  8.21     &   0.140$\pm$0.001     & -5.13     &     0.22$\pm$0.04    \\
HD42719     & 0.657     &  7.51     &   0.138$\pm$0.001     & -5.16     &     0.24$\pm$0.04    \\
HD42936     & 0.900     &  9.10     &   0.157$\pm$0.001     & -5.09     &     0.18$\pm$0.09    \\
HD43197     & 0.817     &  8.98     &   0.146$\pm$0.001     & -5.12     &     0.37$\pm$0.09    \\
HD43306     & 0.835     &  8.70     &   0.168$\pm$0.001     & -5.02     &     0.10$\pm$0.08    \\
HD44310     & 0.837     &  8.68     &   0.187$\pm$0.001     & -4.94     &    -0.02$\pm$0.08    \\
HD44569     & 0.619     &  7.73     &   0.176$\pm$0.001     & -4.90     &    -0.21$\pm$0.03    \\
HD45133     & 0.741     &  8.41     &   0.147$\pm$0.001     & -5.10     &     0.30$\pm$0.06    \\
HD46894     & 0.778     &  7.92     &   0.143$\pm$0.001     & -5.13     &     0.02$\pm$0.06    \\
HD47186     & 0.714     &  7.63     &   0.163$\pm$0.001     & -5.00     &     0.21$\pm$0.06    \\
HD47252     & 0.781     &  8.27     &   0.192$\pm$0.002     & -4.90     &    -0.45$\pm$0.06    \\
HD48265     & 0.747     &  8.05     &  0.128$\pm$0.001     & -5.24     &     0.29$\pm$0.05    \\
HD48611     & 0.742     &  8.71     &  0.228$\pm$0.002     & -4.76     &    -0.32$\pm$0.05    \\
HD49035     & 0.755     &  8.53     &  0.225$\pm$0.002     & -4.77     &     0.29$\pm$0.06    \\
HD49866     & 0.663     &  8.03     &  0.144$\pm$0.001     & -5.11     &     0.10$\pm$0.09    \\
HD50652     & 0.691     &  8.65     &  0.152$\pm$0.001     & -5.06     &     0.24$\pm$0.06    \\
HD52756     & 0.899     &  8.47     &  0.246$\pm$0.002     & -4.83     &     0.09$\pm$0.09    \\
HD55524     & 0.722     &  9.42     &  0.148$\pm$0.001     & -5.09     &     0.24$\pm$0.06    \\
HD56259     & 0.760     &  8.78     &  0.145$\pm$0.001     & -5.12     &     0.14$\pm$0.07    \\
HD56413     & 0.754     &  8.91     &  0.169$\pm$0.001     & -4.98     &     0.24$\pm$0.06    \\
HD56662     & 0.604     &  7.67     &  0.164$\pm$0.001     & -4.96     &    -0.16$\pm$0.03    \\
HD56957     & 0.701     &  7.57     &  0.138$\pm$0.001     & -5.16     &     0.26$\pm$0.05    \\
HD58111     & 0.875     &  8.83     &  0.211$\pm$0.002     & -4.90     &    -0.04$\pm$0.08    \\
HD59100     & 0.634     &  8.17     &  0.172$\pm$0.001     & -4.92     &    -0.15$\pm$0.04    \\
HD61033     & 0.724     &  7.59     &  0.401$\pm$0.003     & -4.40     &    -0.32$\pm$0.06    \\
HD61475     & 0.853     &  8.79     &  0.380$\pm$0.003     & -4.56     &     0.22$\pm$0.07    \\
HD62128     & 0.717     &  7.53     &  0.146$\pm$0.001     & -5.10     &     0.30$\pm$0.07    \\
HD62549     & 0.612     &  7.72     &  0.168$\pm$0.001     & -4.94     &     0.04$\pm$0.06    \\
HD63765     & 0.745     &  8.10     &  0.226$\pm$0.002     & -4.77     &    -0.13$\pm$0.06    \\
HD64010     & 0.815     &  9.26     &  0.418$\pm$0.004     & -4.47     &    -0.19$\pm$0.05    \\
HD65562     & 0.869     &  8.81     &  0.213$\pm$0.002     & -4.89     &    -0.17$\pm$0.08    \\
HD66221     & 0.721     &  8.06     &  0.161$\pm$0.001     & -5.01     &     0.16$\pm$0.06    \\
HD66340     & 0.831     &  9.25     &  0.201$\pm$0.002     & -4.89     &     0.04$\pm$0.07    \\
HD66428     & 0.715     &  8.25     &  0.152$\pm$0.001     & -5.06     &     0.25$\pm$0.05    \\
HD66653     & 0.655     &  7.52     &  0.210$\pm$0.002     & -4.76     &     0.15$\pm$0.06    \\
HD67200     & 0.595     &  7.71     &  0.151$\pm$0.001     & -5.04     &     0.30$\pm$0.04    \\
HD68475     & 0.887     &  8.78     &  0.324$\pm$0.003     & -4.67     &    -0.07$\pm$0.09    \\
HD68785     & 0.619     &  8.19     &  0.196$\pm$0.002     & -4.80     &    -0.21$\pm$0.03    \\
HD68916     & 0.820     &  9.39     &  0.211$\pm$0.002     & -4.86     &    -0.34$\pm$0.06    \\
HD69611     & 0.584     &  7.74     &  0.165$\pm$0.001     & -4.94     &    -0.61$\pm$0.07    \\
HD69721     & 0.826     &  8.80     &  0.167$\pm$0.001     & -5.02     &     0.15$\pm$0.09    \\
HD70081     & 0.660     &  8.92     &  0.147$\pm$0.001     & -5.09     &     0.16$\pm$0.05    \\
HD71251     & 0.851     &  9.19     &  0.371$\pm$0.003     & -4.57     &     0.01$\pm$0.07    \\
HD73536     & 0.698     &  8.19     &  0.164$\pm$0.001     & -4.99     &    -0.06$\pm$0.05    \\
HD75288     & 0.673     &  8.51     &  0.159$\pm$0.001     & -5.01     &     0.09$\pm$0.05    \\
HD75519     & 0.651     &  7.83     &  0.401$\pm$0.003     & -4.35     &     0.13$\pm$0.05    \\
HD76849     & 0.843     &  9.06     &  0.164$\pm$0.001     & -5.04     &     0.05$\pm$0.07    \\
HD77134     & 0.537     &  7.80     &  0.179$\pm$0.001     & -4.84     &    -0.28$\pm$0.01    \\
HD78130     & 0.674     &  8.66     &  0.283$\pm$0.002     & -4.57     &     0.17$\pm$0.04    \\
HD78286     & 0.682     &  8.59     &  0.148$\pm$0.001     & -5.09     &     0.25$\pm$0.05    \\
HD78663     & 0.767     &  8.62     &  0.184$\pm$0.001     & -4.92     &    -0.46$\pm$0.06    \\
HD78747     & 0.575     &  7.72     &  0.170$\pm$0.001     & -4.91     &    -0.69$\pm$0.01    \\
HD81044     & 0.801     &  8.85     &  0.377$\pm$0.003     & -4.51     &    -0.41$\pm$0.07    \\
HD82977     & 0.790     &  9.08     &  0.305$\pm$0.003     & -4.61     &    -0.05$\pm$0.05    \\
HD83517     & 0.664     &  7.74     &  0.172$\pm$0.001     & -4.93     &    -0.23$\pm$0.06    \\
HD84501     & 0.663     &  8.26     &  0.144$\pm$0.001     & -5.12     &     0.09$\pm$0.04    \\
HD84742     & 0.558     &  7.93     &  0.168$\pm$0.001     & -4.91     &    -0.16$\pm$0.43    \\
HD85228     & 0.894     &  7.91     &  0.230$\pm$0.002     & -4.86     &    -0.03$\pm$0.09    \\
HD85404     & 0.591     &  7.78     &  0.152$\pm$0.001     & -5.03     &     0.01$\pm$0.07    \\
HD86006     & 0.708     &  8.17     &  0.155$\pm$0.001     & -5.04     &     0.21$\pm$0.05    \\
HD86171     & 0.746     &  8.53     &  0.260$\pm$0.002     & -4.67     &    -0.17$\pm$0.06    \\
HD87007     & 0.840     &  8.82     &  0.156$\pm$0.001     & -5.07     &     0.16$\pm$0.08    \\
HD87978     & 0.687     &  8.15     &  0.362$\pm$0.003     & -4.43     &     0.02$\pm$0.11    \\
HD88261     & 0.590     &  8.07     &  0.174$\pm$0.001     & -4.89     &    -0.47$\pm$0.21    \\
HD88864     & 0.637     &  7.89     &  0.172$\pm$0.001     & -4.92     &    -0.00$\pm$0.07    \\
HD90028     & 0.656     &  8.32     &  0.150$\pm$0.001     & -5.07     &     0.16$\pm$0.04    \\

\hline
\end{tabular}
\medskip
\end{table*}

\begin{table*}
\center
\begin{tabular}{cccccc}
\hline

HD90520     & 0.643     &  7.51     &  0.161$\pm$0.001     & -4.99     &     0.17$\pm$0.14    \\
HD90722     & 0.724     &  7.88     &  0.146$\pm$0.001     & -5.10     &     0.28$\pm$0.06    \\
HD90812     & 0.813     &  8.86     &  0.182$\pm$0.001     & -4.95     &    -0.33$\pm$0.06    \\
HD91682     & 0.696     &  8.51     &  0.153$\pm$0.001     & -5.06     &     0.16$\pm$0.05    \\
HD92213     & 0.802     &  9.41     &  0.192$\pm$0.002     & -4.91     &    -0.43$\pm$0.07    \\
HD93528     & 0.831     &  8.39     &  0.514$\pm$0.004     & -4.38     &    -0.11$\pm$0.03    \\
HD93626     & 0.536     &  7.88     &  0.177$\pm$0.001     & -4.86     &     0.06$\pm$0.02    \\
HD93849     & 0.559     &  7.86     &  0.147$\pm$0.001     & -5.06     &     0.27$\pm$0.03    \\
HD93932     & 0.615     &  7.53     &  0.154$\pm$0.001     & -5.02     &     0.10$\pm$0.05    \\
HD94151     & 0.718     &  7.84     &  0.184$\pm$0.001     & -4.90     &     0.05$\pm$0.05    \\
HD94268     & 0.557     &  7.63     &  0.175$\pm$0.001     & -4.88     &    -0.39$\pm$0.02    \\
HD94270     & 0.579     &  7.76     &  0.163$\pm$0.001     & -4.95     &     0.00$\pm$0.03    \\
HD94482     & 0.562     &  8.13     &  0.157$\pm$0.001     & -4.99     &    -0.12$\pm$0.01    \\
HD94690     & 0.710     &  8.22     &  0.158$\pm$0.001     & -5.02     &     0.11$\pm$0.05    \\
HD95136     & 0.660     &  8.65     &  0.157$\pm$0.001     & -5.02     &     0.17$\pm$0.04    \\
HD95338     & 0.878     &  8.62     &  0.185$\pm$0.001     & -4.98     &     0.06$\pm$0.08    \\
HD96020     & 0.710     &  8.80     &  0.151$\pm$0.001     & -5.07     &     0.20$\pm$0.06    \\
HD96494     & 0.809     &  8.87     &  0.344$\pm$0.003     & -4.57     &     0.14$\pm$0.06    \\
HD96941     & 0.736     &  8.69     &  0.189$\pm$0.002     & -4.89     &    -0.25$\pm$0.05    \\
HD97038     & 0.661     &  7.70     &  0.155$\pm$0.001     & -5.03     &     0.15$\pm$0.05    \\
HD97356     & 0.637     &  8.15     &  0.166$\pm$0.001     & -4.95     &     0.01$\pm$0.04    \\
HD97517     & 0.613     &  8.11     &  0.165$\pm$0.001     & -4.95     &    -0.35$\pm$0.02    \\
HD98356     & 0.828     &  8.73     &  0.228$\pm$0.002     & -4.81     &     0.07$\pm$0.07    \\
HD98640     & 0.765     &  8.53     &  0.335$\pm$0.003     & -4.54     &    -0.02$\pm$0.03    \\
HD98649     & 0.658     &  8.00     &  0.169$\pm$0.001     & -4.95     &    -0.06$\pm$0.04    \\
HD99565     & 0.748     &  7.65     &  0.254$\pm$0.002     & -4.69     &    -0.37$\pm$0.01    \\
HD100555     & 0.726     &  8.17     & 0.207$\pm$0.002     & -4.81     &     0.07$\pm$0.06    \\
HD100922     & 0.804     &  8.86     & 0.199$\pm$0.002     & -4.88     &    -0.42$\pm$0.04    \\
HD101093     & 0.561     &  7.61     & 0.185$\pm$0.001     & -4.83     &    -0.12$\pm$0.35    \\
HD101197     & 0.659     &  8.73     & 0.171$\pm$0.001     & -4.93     &     0.19$\pm$0.06    \\
HD101348     & 0.711     &  7.82     & 0.138$\pm$0.001     & -5.17     &     0.23$\pm$0.06    \\
HD102196     & 0.569     &  8.15     & 0.146$\pm$0.001     & -5.07     &     0.13$\pm$0.03    \\
HD102346     & 0.551     &  8.15     & 0.322$\pm$0.003     & -4.42     &     0.25$\pm$0.05    \\
HD102361     & 0.556     &  8.16     & 0.157$\pm$0.001     & -4.98     &     0.12$\pm$0.03    \\
HD103197     & 0.860     &  9.40     & 0.158$\pm$0.001     & -5.07     &    -0.12$\pm$0.04    \\
HD103673     & 0.705     &  8.59     & 0.359$\pm$0.003     & -4.45     &     0.02$\pm$0.06    \\
HD104006     & 0.780     &  8.91     & 0.179$\pm$0.001     & -4.95     &    -0.61$\pm$0.05    \\
HD104263     & 0.753     &  8.23     & 0.169$\pm$0.001     & -4.98     &    -0.04$\pm$0.06    \\
HD104576     & 0.708     &  8.53     & 0.422$\pm$0.004     & -4.36     &    -0.55$\pm$0.05    \\
HD104988     & 0.754     &  8.16     &  0.166$\pm$0.001     & -5.00     &    -0.24$\pm$0.05    \\
HD105750     & 0.739     &  8.96     &  0.157$\pm$0.001     & -5.04     &     0.20$\pm$0.03    \\
HD105837     & 0.570     &  7.52     &  0.185$\pm$0.001     & -4.83     &    -0.34$\pm$0.04    \\
HD105901     & 0.626     &  8.20     &  0.168$\pm$0.001     & -4.94     &     0.03$\pm$0.05    \\
HD106275     & 0.892     &  8.63     &  0.207$\pm$0.002     & -4.92     &    -0.14$\pm$0.02    \\
HD106937     & 0.762     &  8.50     &  0.137$\pm$0.001     & -5.17     &     0.18$\pm$0.07    \\
HD107181     & 0.737     &  8.43     &  0.140$\pm$0.001     & -5.15     &     0.26$\pm$0.07    \\
HD107859     & 0.840     &  9.42     &  0.182$\pm$0.001     & -4.96     &    -0.39$\pm$0.05    \\
HD108341     & 0.920     &  9.38     &  0.181$\pm$0.001     & -5.02     &    -0.12$\pm$0.05    \\
HD108446     & 0.872     &  8.72     &  0.331$\pm$0.003     & -4.65     &    -0.19$\pm$0.06    \\
HD108523     & 0.710     &  8.31     &  0.164$\pm$0.001     & -4.99     &    -0.00$\pm$0.06    \\
HD108682     & 0.784     &  9.06     &  0.216$\pm$0.002     & -4.82     &    -0.43$\pm$0.04    \\
HD108953     & 0.800     &  9.26     &  0.174$\pm$0.001     & -4.98     &     0.29$\pm$0.07    \\
HD109423     & 0.898     &  8.82     &  0.423$\pm$0.004     & -4.55     &     0.06$\pm$0.01    \\
HD109570     & 0.692     &  8.04     &  0.217$\pm$0.002     & -4.76     &    -0.05$\pm$0.05    \\
HD109908     & 0.583     &  8.34     &  0.294$\pm$0.002     & -4.49     &     0.13$\pm$0.05    \\
HD109930     & 0.894     &  8.52     &  0.173$\pm$0.001     & -5.03     &     0.41$\pm$0.07    \\
HD110255     & 0.882     &  9.34     &  0.191$\pm$0.002     & -4.96     &    -0.14$\pm$0.04    \\
HD110605     & 0.730     &  9.13     &  0.158$\pm$0.001     & -5.03     &     0.23$\pm$0.07    \\
HD110652     & 0.819     &  9.47     &  0.171$\pm$0.001     & -4.99     &    -0.39$\pm$0.05    \\
HD110979     & 0.654     &  8.05     &  0.163$\pm$0.001     & -4.98     &     0.09$\pm$0.06    \\
HD111096     & 0.703     &  7.69     &  0.153$\pm$0.001     & -5.05     &     0.25$\pm$0.06    \\
HD111431     & 0.627     &  8.01     &  0.144$\pm$0.001     & -5.11     &     0.04$\pm$0.05    \\
HD111777     & 0.616     &  8.48     &  0.173$\pm$0.001     & -4.91     &    -0.44$\pm$0.03    \\
HD112037     & 0.658     &  8.53     &  0.185$\pm$0.001     & -4.87     &    -0.11$\pm$0.05    \\
HD112121     & 0.728     &  8.93     &  0.149$\pm$0.001     & -5.09     &     0.13$\pm$0.06    \\
HD112758     & 0.769     &  7.54     &  0.176$\pm$0.001     & -4.96     &    -0.50$\pm$0.04    \\
HD112863     & 0.779     &  8.70     &  0.407$\pm$0.003     & -4.45     &    -0.30$\pm$0.03    \\
HD113933     & 0.746     &  8.75     &  0.158$\pm$0.001     & -5.03     &     0.01$\pm$0.06    \\
HD117579     & 0.721     &  8.87     &  0.169$\pm$0.001     & -4.97     &    -0.14$\pm$0.05    \\
HD119586     & 0.870     &  9.12     &  0.171$\pm$0.001     & -5.02     &    -0.75$\pm$0.01    \\
HD119607     & 0.876     &  9.21     &  0.409$\pm$0.003     & -4.54     &    -0.16$\pm$0.05    \\

\hline
\end{tabular}
\medskip
\end{table*}

\begin{table*}
\center
\begin{tabular}{cccccc}
\hline

HD120067     & 0.797     &  8.70     &  0.328$\pm$0.003     & -4.58     &    -0.21$\pm$0.05    \\
HD123433     & 0.875     &  9.25     &  0.182$\pm$0.001     & -4.99     &    -0.17$\pm$0.06    \\
HD124595     & 0.610     &  8.76     &  0.152$\pm$0.001     & -5.04     &     0.06$\pm$0.05    \\
HD126535     & 0.840     &  8.86     &  0.479$\pm$0.004     & -4.43     &     0.13$\pm$0.07    \\
HD126828     & 0.533     &  8.25     &  0.142$\pm$0.001     & -5.10     &    -0.06$\pm$0.04    \\
HD127423     & 0.569     &  8.53     &  0.239$\pm$0.002     & -4.63     &     0.13$\pm$0.05    \\
HD128356     & 0.685     &  8.29     &  0.218$\pm$0.002     & -4.75     &    -0.78$\pm$0.03    \\
HD128985     & 0.833     &  9.24     &  0.172$\pm$0.001     & -5.00     &    -0.37$\pm$0.04    \\
HD129445     & 0.756     &  8.80     &  0.175$\pm$0.001     & -4.95     &     0.27$\pm$0.08    \\
HD143120     & 0.780     &  7.52     &  0.125$\pm$0.001     & -5.26     &     0.23$\pm$0.07    \\
HD143361     & 0.773     &  9.20     &  0.136$\pm$0.001     & -5.18     &     0.06$\pm$0.07    \\
HD144550     & 0.682     &  8.61     &  0.137$\pm$0.001     & -5.17     &     0.22$\pm$0.06    \\
HD144848     & 0.648     &  8.66     &  0.140$\pm$0.001     & -5.15     &     0.28$\pm$0.07    \\
HD144899     & 0.660     &  8.97     &  0.132$\pm$0.001     & -5.22     &     0.36$\pm$0.20    \\
HD146233     & 0.652     &  5.49     &  0.167$\pm$0.001     & -4.96     &     0.30$\pm$0.03    \\
HD146434     & 0.700     &  8.14     &  0.154$\pm$0.001     & -5.05     &     0.09$\pm$0.06    \\
HD146856     & 0.870     &  9.40     &  0.332$\pm$0.003     & -4.64     &    -0.23$\pm$0.06    \\
HD146856     & 0.870     &  9.40     &  0.357$\pm$0.003     & -4.61     &    -0.23$\pm$0.07    \\
HD147092     & 0.798     &  9.41     &  0.127$\pm$0.001     & -5.23     &    -0.32$\pm$0.02    \\
HD148156     & 0.560     &  7.69     &  0.145$\pm$0.001     & -5.08     &     0.22$\pm$0.15    \\
HD148319     & 0.645     &  8.58     &  0.284$\pm$0.002     & -4.55     &     0.28$\pm$0.06    \\
HD148577     & 0.664     &  7.93     &  0.137$\pm$0.001     & -5.17     &     0.06$\pm$0.11    \\
HD148577     & 0.664     &  7.93     &  0.139$\pm$0.001     & -5.15     &     0.08$\pm$0.32    \\
HD149189     & 0.653     &  8.56     &  0.133$\pm$0.001     & -5.21     &     0.27$\pm$0.16    \\
HD149194     & 0.559     &  8.68     &  0.153$\pm$0.001     & -5.01     &     0.13$\pm$0.04    \\
HD149616     & 0.561     &  8.21     &  0.152$\pm$0.001     & -5.02     &     0.08$\pm$0.01    \\
HD149724     & 0.783     &  7.85     &  0.151$\pm$0.001     & -5.08     &     0.30$\pm$0.07    \\
HD149782     & 0.759     &  9.16     &  0.188$\pm$0.002     & -4.90     &     0.28$\pm$0.11    \\
HD149782     & 0.759     &  9.16     &  0.189$\pm$0.002     & -4.90     &     0.25$\pm$0.10    \\
HD150437     & 0.683     &  7.84     &  0.127$\pm$0.001     & -5.27     &     0.39$\pm$0.09    \\
HD150936     & 0.740     &  8.76     &  0.172$\pm$0.001     & -4.96     &     0.16$\pm$0.07    \\
HD152079     & 0.711     &  9.18     &  0.152$\pm$0.001     & -5.06     &     0.32$\pm$0.09    \\
HD154221     & 0.640     &  8.60     &  0.157$\pm$0.001     & -5.01     &     0.23$\pm$0.06    \\
HD154672     & 0.713     &  8.21     &  0.138$\pm$0.001     & -5.17     &     0.21$\pm$0.07    \\
HD157691     & 0.642     &  8.37     &  0.161$\pm$0.001     & -4.99     &    -0.30$\pm$0.05    \\
HD157798     & 0.684     &  8.15     &  0.135$\pm$0.001     & -5.19     &     0.00$\pm$0.11    \\
HD158469     & 0.556     &  7.93     &  0.136$\pm$0.001     & -5.16     &     0.11$\pm$0.04    \\
HD162816     & 0.630     &  8.48     &  0.139$\pm$0.001     & -5.15     &     0.08$\pm$0.07    \\
HD165155     & 1.018     &  9.36     &  0.151$\pm$0.001     & -5.20     &     0.10$\pm$0.08    \\
HD165204     & 0.770     &  9.14     &  0.134$\pm$0.001     & -5.19     &     0.30$\pm$0.07    \\
HD165920     & 0.842     &  7.91     &  0.151$\pm$0.001     & -5.09     &     0.19$\pm$0.08    \\
HD166101     & 0.793     &  8.83     &  0.410$\pm$0.003     & -4.46     &    -0.22$\pm$0.06    \\
HD166745     & 0.763     &  8.49     &  0.137$\pm$0.001     & -5.17     &     0.23$\pm$0.08    \\
HD170706     & 0.789     &  9.48     &  0.134$\pm$0.001     & -5.19     &     0.24$\pm$0.09    \\
HD173859     & 0.737     &  8.82     &  0.372$\pm$0.003     & -4.46     &     0.11$\pm$0.06    \\
HD173872     & 0.889     &  8.45     &  0.180$\pm$0.001     & -5.00     &    -0.54$\pm$0.03    \\
HD174541     & 0.631     &  8.59     &  0.143$\pm$0.001     & -5.11     &    -0.13$\pm$0.05    \\
HD175129     & 0.622     &  8.47     &  0.308$\pm$0.003     & -4.48     &    -0.17$\pm$0.10    \\
HD175167     & 0.751     &  8.01     &  0.134$\pm$0.001     & -5.19     &    -0.17$\pm$0.14    \\
HD176157     & 0.829     &  8.39     &  0.204$\pm$0.002     & -4.88     &    -0.20$\pm$0.06    \\
HD178340     & 0.757     &  8.17     &  0.142$\pm$0.001     & -5.13     &     0.24$\pm$0.08    \\
HD178340     & 0.757     &  8.17     &  0.144$\pm$0.001     & -5.12     &     0.32$\pm$0.07    \\
HD178787     & 0.866     &  9.16     &  0.261$\pm$0.002     & -4.77     &    -0.12$\pm$0.05    \\
HD185181     & 0.872     &  9.11     &  0.891$\pm$0.008     & -4.17     &     0.09$\pm$0.09    \\
HD185679     & 0.686     &  8.72     &  0.160$\pm$0.001     & -5.00     &     0.14$\pm$0.07    \\
HD186194     & 0.697     &  8.64     &  0.145$\pm$0.001     & -5.11     &     0.18$\pm$0.07    \\
HD186265     & 0.790     &  9.26     &  0.139$\pm$0.001     & -5.15     &     0.32$\pm$0.09    \\
HD187620     & 0.692     &  8.43     &  0.147$\pm$0.001     & -5.09     &    -0.03$\pm$0.06    \\
HD187694     & 0.746     &  9.36     &  0.161$\pm$0.001     & -5.02     &    -0.21$\pm$0.08    \\
HD188298     & 0.657     &  8.46     &  0.240$\pm$0.002     & -4.67     &     0.13$\pm$0.06    \\
HD188581     & 0.614     &  8.35     &  0.307$\pm$0.003     & -4.48     &    -0.21$\pm$0.04    \\
HD188903     & 0.565     &  8.27     &  0.142$\pm$0.001     & -5.10     &    -0.51$\pm$0.63    \\
HD189627     & 0.556     &  8.11     &  0.128$\pm$0.001     & -5.26     &     0.37$\pm$0.04    \\
HD190125     & 0.708     &  8.93     &  0.227$\pm$0.002     & -4.74     &     0.22$\pm$0.09    \\
HD190613     & 0.629     &  8.12     &  0.139$\pm$0.001     & -5.15     &     0.13$\pm$0.06    \\
HD190647     & 0.743     &  7.78     &  0.132$\pm$0.001     & -5.21     &     0.23$\pm$0.07    \\
HD191122     & 0.643     &  8.65     &  0.147$\pm$0.001     & -5.09     &     0.37$\pm$0.09    \\
HD191760     & 0.668     &  8.26     &  0.137$\pm$0.001     & -5.17     &     0.29$\pm$0.07    \\
HD192771     & 0.723     &  8.64     &  0.160$\pm$0.001     & -5.02     &    -0.15$\pm$0.05    \\
HD193690     & 0.790     &  9.37     &  0.185$\pm$0.001     & -4.93     &     0.24$\pm$0.08    \\
HD193844     & 0.888     &  9.24     &  0.238$\pm$0.002     & -4.84     &    -0.20$\pm$0.01    \\
HD193995     & 0.716     &  8.51     &  0.132$\pm$0.001     & -5.21     &     0.23$\pm$0.07    \\

\hline
\end{tabular}
\medskip
\end{table*}

\begin{table*}
\center
\begin{tabular}{cccccc}
\hline

HD194490     & 0.637     &  8.99     &  0.137$\pm$0.001     & -5.17     &     0.28$\pm$0.08    \\
HD195145     & 0.738     &  8.66     &  0.161$\pm$0.001     & -5.02     &     0.26$\pm$0.08    \\
HD197499     & 0.576     &  8.62     &  0.149$\pm$0.001     & -5.05     &     0.33$\pm$0.07    \\
HD200869     & 0.851     &  9.38     &  0.140$\pm$0.001     & -5.15     &     0.16$\pm$0.09    \\
HD201378     & 0.625     &  7.62     &  0.302$\pm$0.003     & -4.50     &    -0.14$\pm$0.04    \\
HD201757     & 0.714     &  8.19     &  0.134$\pm$0.001     & -5.20     &     0.15$\pm$0.06    \\
HD203384     & 0.762     &  8.03     &  0.163$\pm$0.001     & -5.01     &     0.22$\pm$0.07    \\
HD204807     & 0.717     &  8.28     &  0.163$\pm$0.001     & -5.00     &    -0.20$\pm$0.06    \\
HD204941     & 0.878     &  8.45     &  0.188$\pm$0.002     & -4.97     &    -0.25$\pm$0.06    \\
HD205855     & 0.844     &  8.62     &  0.169$\pm$0.001     & -5.02     &    -0.17$\pm$0.07    \\
HD205871     & 0.641     &  8.39     &  0.158$\pm$0.001     & -5.01     &    -0.25$\pm$0.05    \\
HD206116     & 0.576     &  7.61     &  0.133$\pm$0.001     & -5.20     &     0.25$\pm$0.33    \\
HD206172     & 0.672     &  8.54     &  0.177$\pm$0.001     & -4.91     &    -0.08$\pm$0.06    \\
HD206683     & 0.657     &  8.32     &  0.148$\pm$0.001     & -5.08     &     0.33$\pm$0.06    \\
HD207832     & 0.691     &  8.78     &  0.207$\pm$0.002     & -4.80     &     0.16$\pm$0.07    \\
HD208672     & 0.641     &  8.30     &  0.216$\pm$0.002     & -4.73     &     0.18$\pm$0.07    \\
HD209713     & 0.754     &  9.07     &  0.289$\pm$0.002     & -4.62     &     0.03$\pm$0.08    \\
HD212521     & 0.770     &  8.63     &  0.419$\pm$0.004     & -4.42     &     0.02$\pm$0.06    \\
HD216777     & 0.659     &  8.01     &  0.166$\pm$0.001     & -4.96     &    -0.21$\pm$0.05    \\
HD218249     & 0.879     &  9.33     &  0.196$\pm$0.002     & -4.94     &    -0.73$\pm$0.01    \\
HD218249     & 0.879     &  9.33     &  0.203$\pm$0.002     & -4.92     &    -0.72$\pm$0.01    \\
HD218750     & 0.880     &  9.25     &  0.163$\pm$0.001     & -5.06     &    -0.17$\pm$0.07    \\
HD218960     & 0.667     &  9.08     &  0.138$\pm$0.001     & -5.16     &     0.21$\pm$0.07    \\
HD219011     & 0.727     &  8.90     &  0.135$\pm$0.001     & -5.19     &     0.26$\pm$0.08    \\
HD219556     & 0.759     &  9.08     &  0.145$\pm$0.001     & -5.11     &     0.04$\pm$0.08    \\
HD219556     & 0.759     &  9.08     &  0.145$\pm$0.001     & -5.11     &     0.05$\pm$0.08    \\
HD220256     & 0.853     &  8.56     &  0.166$\pm$0.001     & -5.03     &    -0.25$\pm$0.06    \\
HD220981     & 0.740     &  8.75     &  0.142$\pm$0.001     & -5.13     &     0.18$\pm$0.07    \\
HD221343     & 0.657     &  8.37     &  0.232$\pm$0.002     & -4.69     &     0.16$\pm$0.07    \\
HD221575     & 0.890     &  8.81     &  0.523$\pm$0.004     & -4.44     &     0.28$\pm$0.03    \\
HD221954     & 0.750     &  8.72     &  0.139$\pm$0.001     & -5.16     &     0.27$\pm$0.07    \\
HD222013     & 0.809     &  9.22     &  0.161$\pm$0.001     & -5.04     &     0.05$\pm$0.08    \\
HD222342     & 0.654     &  8.87     &  0.180$\pm$0.001     & -4.89     &     0.11$\pm$0.07    \\
HD222910     & 0.765     &  9.24     &  0.132$\pm$0.001     & -5.21     &     0.12$\pm$0.09    \\
HD223078     & 0.882     &  8.93     &  0.210$\pm$0.002     & -4.91     &    -0.73$\pm$0.01    \\
HD223315     & 0.742     &  8.78     &  0.208$\pm$0.002     & -4.82     &     0.25$\pm$0.08    \\
HD223713     & 0.794     &  9.43     &  0.162$\pm$0.001     & -5.03     &    -0.43$\pm$0.05    \\
HD224065     & 0.680     &  8.86     &  0.180$\pm$0.001     & -4.90     &    -0.18$\pm$0.07    \\
HD224433     & 0.748     &  8.84     &  0.165$\pm$0.001     & -5.00     &     0.12$\pm$0.07    \\
HD224538     & 0.581     &  8.06     &  0.148$\pm$0.001     & -5.06     &     0.39$\pm$0.06    \\
HD224828     & 0.642     &  8.57     &  0.168$\pm$0.001     & -4.95     &    -0.23$\pm$0.07    \\
HD224908     & 0.620     &  8.64     &  0.186$\pm$0.001     & -4.84     &     0.00$\pm$0.06    \\    
HD302554     & 0.705     &  8.83     &  0.332$\pm$0.003     & -4.50     &   -0.02$\pm$0.06        \\
HIP37727     & 0.700     &  7.55     &  0.288$\pm$0.003     & -4.58     &   -0.57$\pm$0.06        \\

\hline
\end{tabular}
\medskip
\end{table*}%

\bibliographystyle{aa}
\bibliography{refs}

\begin{thebibliography}{57}
\expandafter\ifx\csname natexlab\endcsname\relax\def\natexlab#1{#1}\fi

\bibitem[{{Alonso} {et~al.}(1996){Alonso}, {Arribas}, \&
  {Martinez-Roger}}]{alonso96}
{Alonso}, A., {Arribas}, S., \& {Martinez-Roger}, C. 1996, A\&A, 313, 873

\bibitem[{{Asplund} {et~al.}(2000){Asplund}, {Nordlund}, {Trampedach}, {Allende
  Prieto}, \& {Stein}}]{asplund00}
{Asplund}, M., {Nordlund}, {\AA}., {Trampedach}, R., {Allende Prieto}, C., \&
  {Stein}, R.~F. 2000, A\&A, 359, 729

\bibitem[{{Baliunas} \& {Jastrow}(1990)}]{baliunas90}
{Baliunas}, S. \& {Jastrow}, R. 1990, Nature, 348, 520

\bibitem[{{Baliunas} {et~al.}(1995){Baliunas}, {Donahue}, {Soon}, {Gilliland},
  \& {Soderblom}}]{baliunas}
{Baliunas}, S.~L., {Donahue}, R.~A., {Soon}, W., {Gilliland}, R., \&
  {Soderblom}, D.~R. 1995, Bulletin of the American Astronomical Society, 27,
  839

\bibitem[{{Bigot} \& {Th{\'e}venin}(2006)}]{bigot06}
{Bigot}, L. \& {Th{\'e}venin}, F. 2006, MNRAS, 372, 609

\bibitem[{{Blackwell} \& {Lynas-Gray}(1994)}]{blackwell94}
{Blackwell}, D.~E. \& {Lynas-Gray}, A.~E. 1994, A\&A, 282, 899

\bibitem[{{Bond} {et~al.}(2006){Bond}, {Tinney}, {Butler}, {Jones}, {Marcy},
  {Penny}, \& {Carter}}]{bond06}
{Bond}, J.~C., {Tinney}, C.~G., {Butler}, R.~P., {et~al.} 2006, MNRAS, 370, 163

\bibitem[{{Carpenter}(2001)}]{carpenter01}
{Carpenter}, J.~M. 2001, AJ, 121, 2851

\bibitem[{{Duncan} {et~al.}(1991){Duncan}, {Vaughan}, {Wilson}, {Preston},
  {Frazer}, {Lanning}, {Misch}, {Mueller}, {Soyumer}, {Woodard}, {Baliunas},
  {Noyes}, {Hartmann}, {Porter}, {Zwaan}, {Middelkoop}, {Rutten}, \&
  {Mihalas}}]{duncan}
{Duncan}, D.~K., {Vaughan}, A.~H., {Wilson}, O.~C., {et~al.} 1991, ApJS, 76,
  383

\bibitem[{{Edvardsson} {et~al.}(1993){Edvardsson}, {Andersen}, {Gustafsson},
  {Lambert}, {Nissen}, \& {Tomkin}}]{edvardsson93}
{Edvardsson}, B., {Andersen}, J., {Gustafsson}, B., {et~al.} 1993, A\&A, 275,
  101

\bibitem[{{Fischer} \& {Valenti}(2005)}]{fischer05}
{Fischer}, D.~A. \& {Valenti}, J. 2005, ApJ, 622, 1102

\bibitem[{{Gonzalez}(1997)}]{gonzalez}
{Gonzalez}, G. 1997, MNRAS, 285, 403

\bibitem[{{Gray}(2005)}]{gray05}
{Gray}, D.~F. 2005, {The Observation and Analysis of Stellar Photospheres} (The
  Observation and Analysis of Stellar Photospheres, 3rd Edition, by D.F.~Gray.~
  ISBN
  0521851866.~http://www.cambridge.org/us/catalogue/catalogue.asp?isbn=0521851%
866.~Cambridge, UK: Cambridge University Press, 2005.)

\bibitem[{{Gray} {et~al.}(2006){Gray}, {Corbally}, {Garrison}, {McFadden},
  {Bubar}, {McGahee}, {O'Donoghue}, \& {Knox}}]{gray06}
{Gray}, R.~O., {Corbally}, C.~J., {Garrison}, R.~F., {et~al.} 2006, AJ, 132,
  161

\bibitem[{{Gray} {et~al.}(2003){Gray}, {Corbally}, {Garrison}, {McFadden}, \&
  {Robinson}}]{gray03}
{Gray}, R.~O., {Corbally}, C.~J., {Garrison}, R.~F., {McFadden}, M.~T., \&
  {Robinson}, P.~E. 2003, AJ, 126, 2048

\bibitem[{{Grether} \& {Lineweaver}(2006)}]{grether06}
{Grether}, D. \& {Lineweaver}, C.~H. 2006, ApJ, 640, 1051

\bibitem[{{Gurtovenko} \& {Kostik}(1998)}]{gurtovenko98}
{Gurtovenko}, E.~A. \& {Kostik}, R.~I. 1998, Main Astronomical Observatory NAS
  Ukraine, 3E

\bibitem[{{Hall} {et~al.}(2007){Hall}, {Lockwood}, \& {Skiff}}]{hall07}
{Hall}, J.~C., {Lockwood}, G.~W., \& {Skiff}, B.~A. 2007, AJ, 133, 862

\bibitem[{{Hauck} \& {Mermilliod}(1998)}]{hauck}
{Hauck}, B. \& {Mermilliod}, M. 1998, A\&AS, 129, 431

\bibitem[{{Haywood}(2002)}]{haywood}
{Haywood}, M. 2002, MNRAS, 337, 151

\bibitem[{{Henry} {et~al.}(2002){Henry}, {Donahue}, \& {Baliunas}}]{henry02}
{Henry}, G.~W., {Donahue}, R.~A., \& {Baliunas}, S.~L. 2002, ApJL, 577, L111

\bibitem[{{Henry} {et~al.}(1996){Henry}, {Soderblom}, {Donahue}, \&
  {Baliunas}}]{henry}
{Henry}, T.~J., {Soderblom}, D.~R., {Donahue}, R.~A., \& {Baliunas}, S.~L.
  1996, AJ, 111, 439

\bibitem[{{Jenkins} {et~al.}(2006){Jenkins}, {Jones}, {Tinney}, {Butler},
  {McCarthy}, {Marcy}, {Pinfield}, {Carter}, \& {Penny}}]{jenkins06c}
{Jenkins}, J.~S., {Jones}, H.~R.~A., {Tinney}, C.~G., {et~al.} 2006, MNRAS,
  372, 163

\bibitem[{{Jones} {et~al.}(2002){Jones}, {Butler}, {Marcy}, \&
  {Tinney}}]{jones02b}
{Jones}, H.~R.~A., {Butler}, R.~P., {Marcy}, G.~W., \& {Tinney}, C.~G. 2002,
  MNRAS, 337, 1170

\bibitem[{{Kaufer} {et~al.}(1999){Kaufer}, {Stahl}, {Tubbesing}, {Norregaard},
  {Avila}, {Francois}, {Pasquini}, \& {Pizzella}}]{kaufer99}
{Kaufer}, A., {Stahl}, O., {Tubbesing}, S., {et~al.} 1999, The Messenger, 95, 8

\bibitem[{{Kurucz}(1993)}]{kurucz93}
{Kurucz}, R. 1993, ATLAS9 Stellar Atmosphere Programs and 2 km/s grid.~Kurucz
  CD-ROM No.~13.~ Cambridge, Mass.: Smithsonian Astrophysical Observatory,
  1993., 13

\bibitem[{{Kurucz}(1979)}]{kurucz79}
{Kurucz}, R.~L. 1979, ApJS, 40, 1

\bibitem[{{Kurucz}(1992)}]{kurucz92}
{Kurucz}, R.~L. 1992, Revista Mexicana de Astronomia y Astrofisica, vol.~23,
  23, 45

\bibitem[{{Lockwood} {et~al.}(2007){Lockwood}, {Skiff}, {Henry}, {Henry},
  {Radick}, {Baliunas}, {Donahue}, \& {Soon}}]{lockwood07}
{Lockwood}, G.~W., {Skiff}, B.~A., {Henry}, G.~W., {et~al.} 2007, ApJS, 171,
  260

\bibitem[{{Middelkoop}(1982{\natexlab{a}})}]{middelkoop82}
{Middelkoop}, F. 1982{\natexlab{a}}, A\&A, 107, 31

\bibitem[{{Middelkoop}(1982{\natexlab{b}})}]{middelkoop82b}
{Middelkoop}, F. 1982{\natexlab{b}}, A\&A, 113, 1

\bibitem[{{Moore}(1956)}]{moore56}
{Moore}, C.~E. 1956, Vistas in Astronomy, 2, 1209

\bibitem[{{Nordstr{\"o}m} {et~al.}(2004){Nordstr{\"o}m}, {Mayor}, {Andersen},
  {Holmberg}, {Pont}, {J{\o}rgensen}, {Olsen}, {Udry}, \&
  {Mowlavi}}]{nordstrom04}
{Nordstr{\"o}m}, B., {Mayor}, M., {Andersen}, J., {et~al.} 2004, A\&A, 418, 989

\bibitem[{{Noyes} {et~al.}(1984a){Noyes}, {Hartmann}, {Baliunas}, {Duncan}, \&
  {Vaughan}}]{noyes}
{Noyes}, R.~W., {Hartmann}, L.~W., {Baliunas}, S.~L., {Duncan}, D.~K., \&
  {Vaughan}, A.~H. 1984a, ApJ, 279, 763

\bibitem[{{Olsen}(1984)}]{olsen84}
{Olsen}, E.~H. 1984, A\&AS, 57, 443

\bibitem[{{Pavlenko}(1991)}]{pavlenko91}
{Pavlenko}, Y.~V. 1991, Soviet Astronomy, 35, 212

\bibitem[{{Pavlenko}(2000)}]{pavlenko00}
{Pavlenko}, Y.~V. 2000, Astronomy Reports, 44, 219

\bibitem[{{Pavlenko} {et~al.}(1995){Pavlenko}, {Rebolo}, {Martin}, \& {Garcia
  Lopez}}]{pavlenko95}
{Pavlenko}, Y.~V., {Rebolo}, R., {Martin}, E.~L., \& {Garcia Lopez}, R.~J.
  1995, A\&A, 303, 807

\bibitem[{{Perryman} {et~al.}(1997){Perryman}, {Lindegren}, {Kovalevsky},
  {Hoeg}, {Bastian}, {Bernacca}, {Cr{\' e}z{\' e}}, {Donati}, {Grenon}, {van
  Leeuwen}, {van der Marel}, {Mignard}, {Murray}, {Le Poole}, {Schrijver},
  {Turon}, {Arenou}, {Froeschl{\' e}}, \& {Petersen}}]{perryman}
{Perryman}, M.~A.~C., {Lindegren}, L., {Kovalevsky}, J., {et~al.} 1997, A\&A,
  323, L49

\bibitem[{{Queloz} {et~al.}(2001){Queloz}, {Henry}, {Sivan}, \&
  {Baliunas}}]{queloz}
{Queloz}, D., {Henry}, G.~W., {Sivan}, J.~P., \& {Baliunas}, S.~L. 2001, A\&A,
  379, 279

\bibitem[{{Rieutord} {et~al.}(2001){Rieutord}, {Roudier}, {Ludwig}, {Nordlund},
  \& {Stein}}]{rieutord01}
{Rieutord}, M., {Roudier}, T., {Ludwig}, H.-G., {Nordlund}, {\AA}., \& {Stein},
  R. 2001, A\&A, 377, L14

\bibitem[{{Rutten}(1984)}]{rutten84}
{Rutten}, R.~G.~M. 1984, A\&A, 130, 353

\bibitem[{{Saar} \& {Baliunas}(1992)}]{saar92}
{Saar}, S.~H. \& {Baliunas}, S.~L. 1992, in ASP Conf. Ser. 27: The Solar Cycle,
  ed. K.~L. {Harvey}, 150--167

\bibitem[{{Saar} {et~al.}(1998){Saar}, {Butler}, \& {Marcy}}]{saar98}
{Saar}, S.~H., {Butler}, R.~P., \& {Marcy}, G.~W. 1998, ApJL, 498, L153

\bibitem[{{Santos} {et~al.}(2004){Santos}, {Israelian}, \& {Mayor}}]{santos04}
{Santos}, N.~C., {Israelian}, G., \& {Mayor}, M. 2004, A\&A, 415, 1153

\bibitem[{{Santos} {et~al.}(2003){Santos}, {Israelian}, {Mayor}, {Rebolo}, \&
  {Udry}}]{santos03}
{Santos}, N.~C., {Israelian}, G., {Mayor}, M., {Rebolo}, R., \& {Udry}, S.
  2003, VizieR Online Data Catalog, 339, 80363

\bibitem[{{Santos} {et~al.}(2000){Santos}, {Mayor}, {Naef}, {Pepe}, {Queloz},
  {Udry}, \& {Blecha}}]{santos00}
{Santos}, N.~C., {Mayor}, M., {Naef}, D., {et~al.} 2000, A\&A, 361, 265

\bibitem[{{Schuster} \& {Nissen}(1989)}]{schuster89}
{Schuster}, W.~J. \& {Nissen}, P.~E. 1989, A\&A, 221, 65

\bibitem[{{Smalley} \& {Kupka}(1997)}]{smalley97}
{Smalley}, B. \& {Kupka}, F. 1997, A\&A, 328, 349

\bibitem[{{Smith}(1995)}]{smith95}
{Smith}, R.~C. 1995, {Observational Astrophysics} (Observational Astrophysics,
  by Robert C.~Smith, pp.~467.~ISBN 0521278341.~Cambridge, UK: Cambridge
  University Press, June 1995.)

\bibitem[{{Tinney} {et~al.}(2002){Tinney}, {McCarthy}, {Jones}, {Butler},
  {Carter}, {Marcy}, \& {Penny}}]{tinney02}
{Tinney}, C.~G., {McCarthy}, C., {Jones}, H.~R.~A., {et~al.} 2002, MNRAS, 332,
  759

\bibitem[{{Twarog} {et~al.}(2002){Twarog}, {Anthony-Twarog}, \&
  {Tanner}}]{twarog02}
{Twarog}, B.~A., {Anthony-Twarog}, B.~J., \& {Tanner}, D. 2002, AJ, 123, 2715

\bibitem[{{Valenti} \& {Fischer}(2005)}]{valenti05}
{Valenti}, J.~A. \& {Fischer}, D.~A. 2005, ApJS, 159, 141

\bibitem[{{Wright}(2004{\natexlab{a}})}]{wright04}
{Wright}, J.~T. 2004{\natexlab{a}}, AJ, 128, 1273

\bibitem[{{Wright}(2004{\natexlab{b}})}]{wright04b}
{Wright}, J.~T. 2004{\natexlab{b}}, AJ, 128, 1273

\bibitem[{{Wright}(2005)}]{wright05}
{Wright}, J.~T. 2005, PASP, 117, 657

\bibitem[{{Zhang} {et~al.}(1994){Zhang}, {Soon}, {Baliunas}, {Lockwood},
  {Skiff}, \& {Radick}}]{zhang94}
{Zhang}, Q., {Soon}, W.~H., {Baliunas}, S.~L., {et~al.} 1994, ApJL, 427, L111

\end{thebibliography}






   
  



\end{document}